\documentclass[manuscript]{aastex}
\usepackage{graphicx,natbib,amsmath}
\newcounter{subeqn} \renewcommand{\thesubeqn}{\theequation\alph{subeqn}}%
\makeatletter
\@addtoreset{subeqn}{equation}
\makeatother
\newcommand{\subeqn}{%
  \refstepcounter{subeqn}
  \tag{\thesubeqn}
}

\newcommand{\mps}{m\,s$^{-1}$}
\newcommand{\bvec}[1]{ \mbox{\boldmath$#1$} }
\def\id{{\rm d}}
\def\bcK{\mbox{\boldmath${\cal K}$}}
\def\bcT{\mbox{\boldmath${\cal T}$}}
\def\cT{{\cal T}}
\def\cK{{\cal K}}

\shorttitle{Tomography of plasma flows in the upper solar convection zone}
\shortauthors{Michal \v{S}vanda}

\begin{document}
\title{Tomography of plasma flows in the upper solar convection zone using time--distance inversion combining ridge and phase-speed filtering}
\author{Michal \v{S}vanda}
\affil{Astronomical Institute, Academy of Sciences of the Czech Republic (v. v. i.), Fri\v{c}ova 298, CZ-25165 Ond\v{r}ejov, Czech Republic}
\affil{Astronomical Institute, Charles University in Prague, Faculty of Mathematics and Physics, V Hole\v{s}ovi\v{c}k\'ach 2, CZ-18000 Prague 8, Czech Republic}
\email{michal@astronomie.cz}

\begin{abstract}
The consistency of time--distance inversions for horizontal components of the plasma flow on supergranular scales in the upper solar convection zone is checked by comparing the results derived using two $k$--$\omega$ filtering procedures -- ridge filtering and phase-speed filtering -- commonly used in time--distance helioseismology. It is shown that both approaches result in similar flow estimates when finite-frequency sensitivity kernels are used. It is further demonstrated that the performance of the inversion improves (in terms of simultaneously better averaging kernel and lower noise level) when the two approaches are combined together in one inversion. Using the combined inversion I invert for horizontal flows in the upper 10~Mm of the solar convection zone. The flows connected with supergranulation seem to be coherent only in the upper $\sim5$~Mm depth, deeper down there is a hint on change of convection scales towards structures larger than supergranules.
\end{abstract}

\keywords{Sun: helioseismology -- Sun: interior}

\section{Measurements of plasma flows by time--distance helioseismology}

Helioseismology is the only one method used in solar physics, which allows us to study details of what is going on below the optically thick photosphere. Inverse modelling in local helioseismology became a standard method of solar research in the past few decades. Inversions for local helioseismology were notably successful in investigating the properties of plasma flows in the solar convection zone. Despite the effort, some open questions with regards to convective flows remain. What is the depth structure and nature of supergranules? What are the properties of deep solar convection? The reliable knowledge of plasma behavior in the Sun is particularly important for solar dynamo, a process which forms all phenomena of solar activity. 

The plasma flows in the upper solar convection zone seem to have multi-scale character \citep[see review by][]{2009LRSP....6....2N}. Two distinct scales however exist: granulation \citep[reviewed by][]{1963ARAA...1...19L} and supergranulation \citep[reviewed by][]{lrsp-2010-2}. While the granulation is considered well understood in terms of underlying physics, the supergranulation is not. The surface velocity spectrum \citep{2000SoPh..193..299H} does not indicate presence of distinct either smaller (termed mesogranules) or larger (termed giant cells) convection modes. At least the giant cells are, however, a common feature seen in some global models of solar convection zone \citep[e.g.][]{2008ApJ...673..557M}. Recently, the fundamental problem in understanding the solar convection emerged: the upper limit set by helioseismology on amplitude of convective velocities in the deep solar convection zone \citep{2012PNAS..10911928H} is two orders smaller than what is predicted from global convection simulations. Even lower limits set by dynamical balance to maintain the observed properties of mean plasma flows \citep{2012ApJ...757..128M} do not reconcile with the measurements from helioseismology. It is believed that only helioseismology may resolve this ``convection crisis'' and this motivates my research. 

From all the methods of helioseismology I focus on time--distance helioseismology \citep{1993Natur.362..430D}, a method used to to measure and interpret changes in travel times of seismic waves caused by inhomogeneities in the structure of the Sun, among which plasma flows play an important role. In recent years, time--distance helioseismology has been used to invert for near-surface flows \citep[e.g.][]{2000JApA...21..339G,2000SoPh..192..177D,2004ApJ...603..776Z,2008SoPh..251..381J}, for flows beneath sunspots \citep[e.g.][]{1996Natur.379..235D,2001ApJ...557..384Z,2006ApJ...640..516C,2008SoPh..251..291C,2009SSRv..144..249G}, flows in their vicinity \citep{2000JApA...21..339G}, and many others. 

Some delicate problems emerged over time. For instance, some authors question a possibility to measure the deep plasma flows on supergranular scales with the signal-to-noise ratio larger than one. Although there are results published that ``image'' the plasma flows on supergranular scales at depths of 64~Mm or so \citep[e.g.][]{1996ApJ...461L..55K}, some more recent studies suggest that the noise dominates the signal already at depths of 4--6~Mm \citep[e.g.][]{2007ApJ...668.1189W,2008SoPh..251..381J,Svanda2011}. The argument that there is a cellular pattern clearly visible in the flow maps from large depths is problematic at least. The danger of ``seeing'' cells in the flow map and connecting them to the convection was shown by \cite{2010AA...519A..52M}, who demonstrated that mesogranules have statistical properties of the averaged noise, thus do not necessarily physically exist. To avoid these issues, it is imperative to use helioseismic methods, which provide (1) the tomographic image of the flows, (2) all the details of the averaging kernel, describing the localisation in the Sun, and (3) the level of random noise in the results. The code I have at my disposal fulfills all these requirements.

\section{Spatio-temporal filters for time--distance helioseismology}
\label{sect:processing_from}
The time--distance inversion pipeline consists of many consecutive steps. The first one typically is a filtering of a time series of observations (most often the series of Dopplergrams) to retain only particular modes of oscillations. The filtering is usually executed by means of spatio-temporal filter applied in the Fourier space. 

In practice, the observations form a datacube $\Phi(\bvec{r},t)$, where $\bvec{r}$ is a horizontal position vector and $t$ time. For convenience, this datacube is Fourier transformed to become $\Phi(\bvec{k},\omega)$, where $\bvec{k}$ is a horizontal wave vector and $\omega$ frequency. Then, the Fourier data series is multiplied by the filter $F(\bvec{k}, \omega)$ in order to obtain the new filtered datacube $\Psi(\bvec{k}, \omega)$: 
\begin{equation}
\Psi(\bvec{k}, \omega)=F(\bvec{k}, \omega) \Phi(\bvec{k},\omega).
\end{equation}
The filtered datacube is consequently analysed by time--distance techniques. 

In time--distance helioseismology, two approaches are usually taken to construct the filter $F$. The approach considered traditional is to retain only acoustic ($p$) modes having the same phase speed $v_{\rm ph}$. The \emph{phase-speed filtering} was introduced by \cite{1997SoPh..170...63D} to deal with artifacts in Dopplergram measurements by Michelson Doppler Imager \citep[MDI;][]{1995SoPh..162..129S} prohibiting to measure the wave travel times for distances shorter than some 10~Mm. Physically, the waves with the same phase speed have (under the approximation by the geometrical optics) the same lower turning point, thus roughly travel the same path in the solar interior \citep{1997ApJ...477..475B}. The phase-speed filters can be computed analytically using the formula 
\begin{equation}
F(k,\omega) = \exp[-{(\omega/k-v_{\rm ph})^2}/{(2\delta v_{\rm ph}^2)}],
\end{equation}
where $k=||\bvec{k}||$. The values of the central phase speed $v_{\rm ph}$ and of the width of the bandpass $\delta v_{\rm ph}$ of eleven filters considered standard in helioseismology are given in Table~\ref{tab:phs_filters}. Hereafter, these eleven standard phase-speed filters are denoted as TD1 to TD11 with increasing phase-speed value.

A very different filtering scheme was recently applied in some time--distance studies \citep[e.g.][to name a few not co-authored by the author of this study]{2000JApA...21..339G,2008SoPh..251..381J,2008SoPh..251..267B}. The technique of \emph{ridge filtering} became common in geoseismology \citep[e.g.][]{1975GeoRL...2...60N,1997GeoJI.131..209H} and employs filters separating the modes of oscillations having the same radial order. Modes having the same radial order form a ``ridge'' in the power spectrum of oscillations. The ridge filters used in this study are constructed in accordance with the corresponding module of the kernel code (written by Aaron Birch): for each wave number $k$, the filter gains the following values:
\begin{equation}
F(k,\omega)\left\{ \begin{array}{l}
=0 \quad \mbox{for} \ \omega < \omega_n - \frac{3}{4}\left(\omega_n-\omega_{n-1} \right) \ \mbox{or}\ \omega > \omega_n + \frac{3}{4}\left(\omega_{n+1}-\omega_{n} \right) \\   
=1 \quad \mbox{for} \ \omega \in \left(\omega_n - \frac{1}{4}(\omega_n-\omega_{n-1}), \omega_n + \frac{1}{4}(\omega_{n+1}-\omega_{n}) \right)\\
=\frac{1}{2} - \frac{1}{2} \cos{\pi\frac{\omega-\frac{1}{4}\omega_n-\frac{3}{4}\omega_{n-1}}{\frac{1}{2}(\omega_n-\omega_{n-1})}}  \ \mbox{for} \ \omega \in \langle\omega_n - \frac{3}{4}\left(\omega_n-\omega_{n-1} \right), \omega_n - \frac{1}{4}\left(\omega_n-\omega_{n-1} \right)\rangle\\
=\frac{1}{2} - \frac{1}{2}\cos{\pi\frac{\omega-\frac{1}{4}\omega_n-\frac{3}{4}\omega_{n+1}}{\frac{1}{2}(\omega_{n+1}-\omega_{n})}} \ \mbox{for} \ \omega \in \langle\omega_n + \frac{1}{4}\left(\omega_{n+1}-\omega_{n} \right), \omega_n + \frac{3}{4}\left(\omega_{n+1}-\omega_{n} \right)\rangle
\end{array}\right.
\end{equation}
where $\omega_n$ is the peak frequency of the ridge to be retained at the given $k$ and $\omega_{n-1}$ and $\omega_{n+1}$ are peak frequencies of the adjacent ridges.  The position of ridges can either be determined from studying the power spectra or from eigenfrequencies of the solar model. For the $f$ mode, the filter is symmetrical in frequency about $\omega_n$ by defining $\omega_{n-1}=\omega_n-(\omega_{n+1}-\omega_n)$. In this study, I use surface gravity mode (the $f-$mode) and first four acoustic ridges $p_1$ to $p_4$.

\begin{figure}
\includegraphics[width=0.5\textwidth]{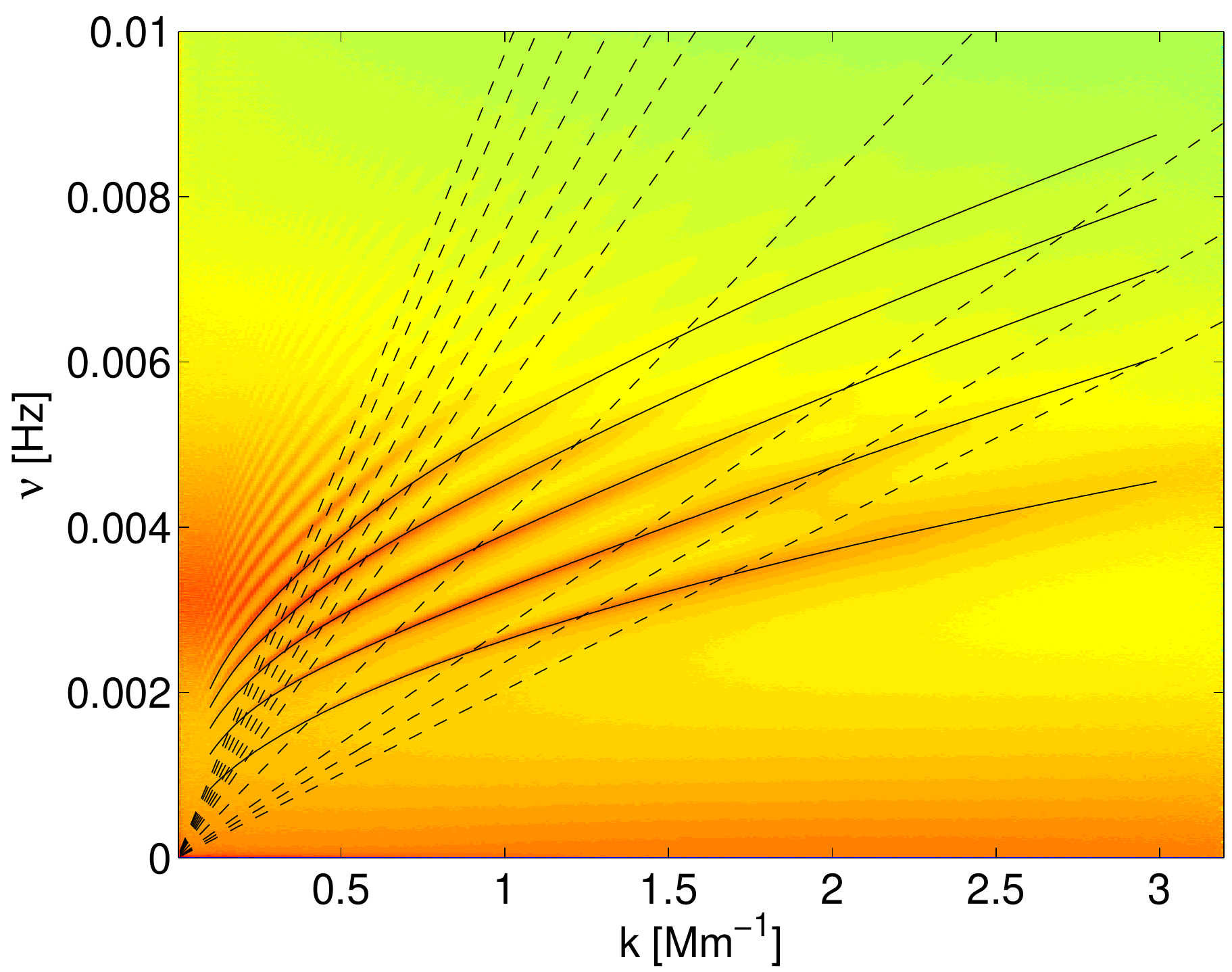}
\caption{Power spectrum of HMI Dopplergram datacube spanning 24 hours with colour table adjusted to increase the visibility of the ridges of oscillations. The solid lines over-plot the resonant eigenfrequencies from Model S \citep{1996Sci...272.1286C}, used to construct the ridge filters ($f$, $p_1$, $p_2$, $p_3$, and $p_4$ in the bottom-up order), the dashed lines over-plot the phase speeds of centres of eleven standard time--distance filters (TD1 to TD11 in the bottom-up order). When phase-speed filters are used, the $f$-mode is filtered out, because it has a different dispersion relation.}
\label{fig:ps_and_filters}
\end{figure}

Positions of all filters used in this study are over-plotted the power spectrum of Dopplergrams measured by Helioseismic and Magnetic Imager \citep[HMI;][]{2012SoPh..275..207S,2012SoPh..275..229S} in Fig.~\ref{fig:ps_and_filters}. Additionally, all datacubes are filtered with the high-pass filter with cut-off frequency of 1.5~mHz to suppress the disturbing effects of surface convection. 

\section{Measurement of travel times}

The wave travel times are measured from the temporal cross-covariance $C(\bvec{r}_1,\bvec{r}_2,t)$ of the filtered signal $\Psi(\bvec{r},t)$ at two points $\bvec{r}_1$ and $\bvec{r}_2$ at the solar surface,
\begin{equation}
C(\bvec{r}_1, \bvec{r}_2 ,t)=\frac{1}{T}  \int\limits_{0}^{T} \id t'\, \Psi(\bvec{r}_1, t') \Psi(\bvec{r}_2,t'+t), 
\label{eq:xcpoint}
\end{equation}
where $T$ is the length of observation. The formalism can be generalized and instead of the point-to-point cross-covariance, other measurement geometries can be introduced. Traditionally, the center-to-annulus geometry is used in time--distance helioseismology, where the filtered signal at a single point $\bvec{r}_2$ is replaced by $\Psi$ averaged over the annulus with radius $\Delta$ around the central point $\bvec{r}_1$. Similarly, the center-to-quadrants geometry is an extension of a center-to-annulus geometry by weighting the surrounding annulus by e.g. cosine and sine of the azimuthal angle. Note that while in the case of ridge filters the user has effectively free choice of annulus radius $\Delta$ (in this study I use the set of distances 7.3~Mm to 29.2~Mm with a step of 1.46~Mm), the range of sensible distances in the case of phase-speed filters is set by the central phase speed and the selections of $\Delta$ used in this study are given in Table~\ref{tab:phs_filters}. The particular combination of the choices in the data analysis (Fourier filter, annulus or quadrant, and the radius of the annulus) are uniquely combined into an index $a$.

For each geometry $a$ the travel time $\tau^a$ is measured at every position $\bvec{r}$ in the field-of-view as \citep[full details in][]{2004ApJ...614..472G} 
\begin{equation}
\tau^a_\pm(\bvec{r})=\int\limits_0^T \id t\, W^a_\pm(t) \left[C^a(\bvec{r},t)-C^{\rm ref}(t) \right],
\label{eq:TTdef}
\end{equation}
where the weighted (by weighting functions $W^a_\pm$) difference between the measured cross-covariance and the reference cross-covariance $C^{\rm ref}$ is integrated over the entire observation time $T$. In this work I choose the reference cross-covariance to be the symmetric component of the spatial average of the quiet Sun cross-covariances.  

The travel time $\tau^a_\pm$ has two branches for positive and negative time lags (indicating the waves travelling from the point to the surrounding annulus or quadrant and the waves travelling from the annulus or quadrant to the central point). For the remainder of this study I focus on difference travel time, thus
\begin{equation}
\delta\tau^a(\bvec{r})=\tau^a_{+}(\bvec{r})-\tau^a_{-}(\bvec{r}).
\end{equation}

The difference travel times are particularly sensitive to plasma flows and can be thus used for their inversion. 

\section{SOLA inversions}
\label{sect:processing_to}
For simplicity, I focus this study to a small patch near the centre of the solar disc, thus only small error is caused \citep{2012SoPh..tmp..306B} by approximating the field of view in Cartesian coordinate system $\bvec{x}=(\bvec{r},z)=(x,y,z)$, where $\bvec{r}$ is (consistently with what was already said earlier) the horizontal position vector and $z$ is the height. Axis $\hat{x}$ points in the solar east--west direction, axis $\hat{y}$ is parallel to the axis of solar rotation. 

In forward modelling, the velocity vector $\bvec{v}$ translates into travel-time deviations $\delta\tau$ via travel-time sensitivity kernel $\bvec{K}^a = (K_x^a, K_y^a, K_z^a)$ assuming the linear relationship
\begin{equation}
\delta\tau^a(\bvec{r}) = \int_{\odot} \bvec{K}^a(\bvec{r'}-\bvec{r},z) \cdot \bvec{v}(\bvec{r'},z)\; \id^2\bvec{r'} \, \id z + n^a(\bvec{r})\ .
\label{eq:traveltimesdef}
\end{equation}
The sensitivity kernels are computed in Born approximation \citep{2007AN....328..228B} and they are consistent with the travel-time measurements by employing exactly same $k$--$\omega$ filters, both observations and sensitivity kernels are processed with the same pixel size, etc. A particular realisation of the random noise $n^a$ is not known, however its covariance matrix can be measured from the large set of travel-time maps \citep[for details see][]{2004ApJ...614..472G}, which is the approach I took. 

The ultimate goal of inverse modelling is to retrieve $\bvec{v}$ from (\ref{eq:traveltimesdef}) when knowing $\delta\tau$ and $\bvec{K}$. Studies have shown that the exact solution is generally not possible and that the inverse problem becomes an optimization problem. One possible solution to the problem is achieved by means of a Subtractive Optimally Localised Averaging \citep[SOLA;][]{1992AA...262L..33P} approach. It aims to construct a spatially bound averaging kernel by linearly combining the set of sensitivity kernels while keeping the error magnification under control. The procedure results in a set of inversion weights $w_a^\alpha$, which are used to linearly combine the travel-time maps in order to get the estimate for the flow component $v_\alpha^{\rm inv}$, where $\alpha=x,y$, using 
\begin{align}
v_\alpha^{\rm inv}(\bvec{r}_0,z_0)&=\sum_a \int w_a^\alpha (\bvec{r'}-\bvec{r}_0;z_0) \tau^a(\bvec{r'}) \id^2\bvec{r'}= \label{eq:inversion}\\
&=\int_\odot \cK^\alpha_\alpha(\bvec{r}-\bvec{r}_0,z;z_0) v_\alpha(\bvec{r},z) \id^2\bvec{r} \id z \refstepcounter{equation}\subeqn \label{eq:inversion1}\\
&+\sum_{\beta \ne \alpha} \int_\odot \cK^\alpha_\beta(\bvec{r}-\bvec{r}_0,z;z_0) v_\beta(\bvec{r},z) \id^2\bvec{r} \id z \subeqn \label{eq:inversion2}\\
&+v_\alpha^{\rm noise}(\bvec{r}_0,z_0) \subeqn \label{eq:inversion3},
\end{align}
where $\beta=x,y,z$. 

The estimate for the flow velocity component is then a combination of a true velocity component smoothed by an averaging kernel $\cK^\alpha_\alpha$ (\ref{eq:inversion1}), a crosstalk from other components (\ref{eq:inversion2}), and a random-noise component $v_\alpha^{\rm noise}$ (\ref{eq:inversion3}), root-mean-square value of which ($\sigma_\alpha$) is to be bound.

In terms of the weights, the component $\cK_\beta^\alpha$ of the averaging kernel is expressed by
\begin{equation}
\cK_\beta^\alpha(\bvec{r}, z;z_0)=\sum_a \int_\odot w_a^\alpha(\bvec{r'};z_0) K_\beta^a(\bvec{r}-\bvec{r'},z) \id^2\bvec{r'}.
\end{equation}

SOLA algorithm searches for the weights that produce the averaging kernel close to a user-supplied \emph{target function} $\bcT_\beta^\alpha(\bvec{r}, z;z_0)$ and that are normalised.
The target function is chosen to have a non-trivial component only in the direction of the inversion, thus
\begin{equation}
\bcT_\beta^\alpha(\bvec{r}, z;z_0)=\cT(\bvec{r}, z;z_0) \delta_\beta^\alpha, 
\end{equation}
where $\delta_\beta^\alpha$ is Kronecker $\delta$. $\cT(\bvec{r}, z;z_0)$ usually has a form of a 3-D Gaussian. 

To find a numerical solution to the problem, a cost function can be constructed, minimising the norm $||\bcK-\bcT||$, the level of random error in the resulting estimate and some other terms using free trade-off parameters. I do not provide the full solution in this paper and rather refer to already published papers containing all mathematical details \citep{fastOLA,Svanda2011}. 

I would only like to make one final point on user's freedom of selection of target functions. It is the usual case that the user wants the inversion to result in the flow estimates that are not difficult to interpret. Thus the signal-to-noise ratio in the results should be reasonably higher than unity and the averaging kernel should have as little side-lobes as possible. The feasibility of the inversion to find an acceptable solution is thus strongly limited by the consistency of the required target function and the set of sensitivity kernels used. This point concerns especially the depth dependence of the target functions. It is thus useful to construct horizontal averages of sensitivity kernels used to get a feeling of what might end up with useful results and what must inevitably fail. Such horizontal averages of sensitivity kernels for each $k$--$\omega$ filter used, averaged additionally over all geometries and distances within the filter and normalised so that the total spatial integral is unity, are displayed in Fig.~\ref{fig:1Dkernels}. 

A curious reader may immediately see that, first, all (!) kernels are highly sensitive at the surface, where also the flows are the largest. Thus the travel-time maps measured using these filters have a dominant contribution from the surface layers and it is up to inversion to deconvolve the weak travel-time signal from deeper layers from such measurements. And second, even with the use of highest phase-speed filter (TD11) the user cannot expect to obtain any meaningful flow inversion from depths deeper than 20~Mm. A ray-theory-based estimate for depth of the lower turning point for this phase speed is around 23~Mm \citep{2013ApJ...762..131B}. To be on the safe side, I limit myself in this study to the depths shallower than 10~Mm. 

\section{Comparison of flows inverted using both filtering approaches}

Each filtered mode of oscillations comprises a rather independent information about the travelling waves. In the end, I would like to combine advantages of both filtering approaches to learn about plasma motions in the near-surface layers of the solar convection zone. To my knowledge, both disjunctive filtering approaches were never combined in one inversion. As a first step, a consistency check must be made. 

This study analyses real observations of the Sun. I used travel-time maps measured from full-disc Dopplergrams observed by HMI in June 2011. The Dopplergrams were tracked and remapped to Postel projection using standard data processing technique. Only the disc-centre region (512$\times$512 pixels) was tracked always for 24 hours to conform with the approximation by the Cartesian coordinate system with a pixel size of 1.46~Mm. The tracking and mapping was done using the code {\sc drms\_tracking} (Schunker \& Burston, unpublished) implemented within German Science Center for SDO at Max-Planck-Institut f\"ur Sonnensystemforschung, Katlenburg-Lindau, Germany. Over the two months interval I restricted only to days with no activity in the field of view: This strong constraint allowed to keep only nine days for a further investigation. 

The sensitivity kernels at my disposal were computed for analysis of MDI full disc data, the predecessor of modern HMI. HMI is a better resolution instrument and thus the power spectra of oscillations of the two are not exactly the same, perhaps due to effect of the different spectral lines used to measure the Doppler shifts. Thus, in accordance with \cite{2013arXiv1302.0790S}, the power spectrum of each tracked HMI Dopplergram datacube is corrected to resemble the power spectrum of full-disc MDI observations. Selected observations underwent consistent processing described in Sections \ref{sect:processing_from} to \ref{sect:processing_to}. 

\begin{figure}
\includegraphics[width=0.5\textwidth]{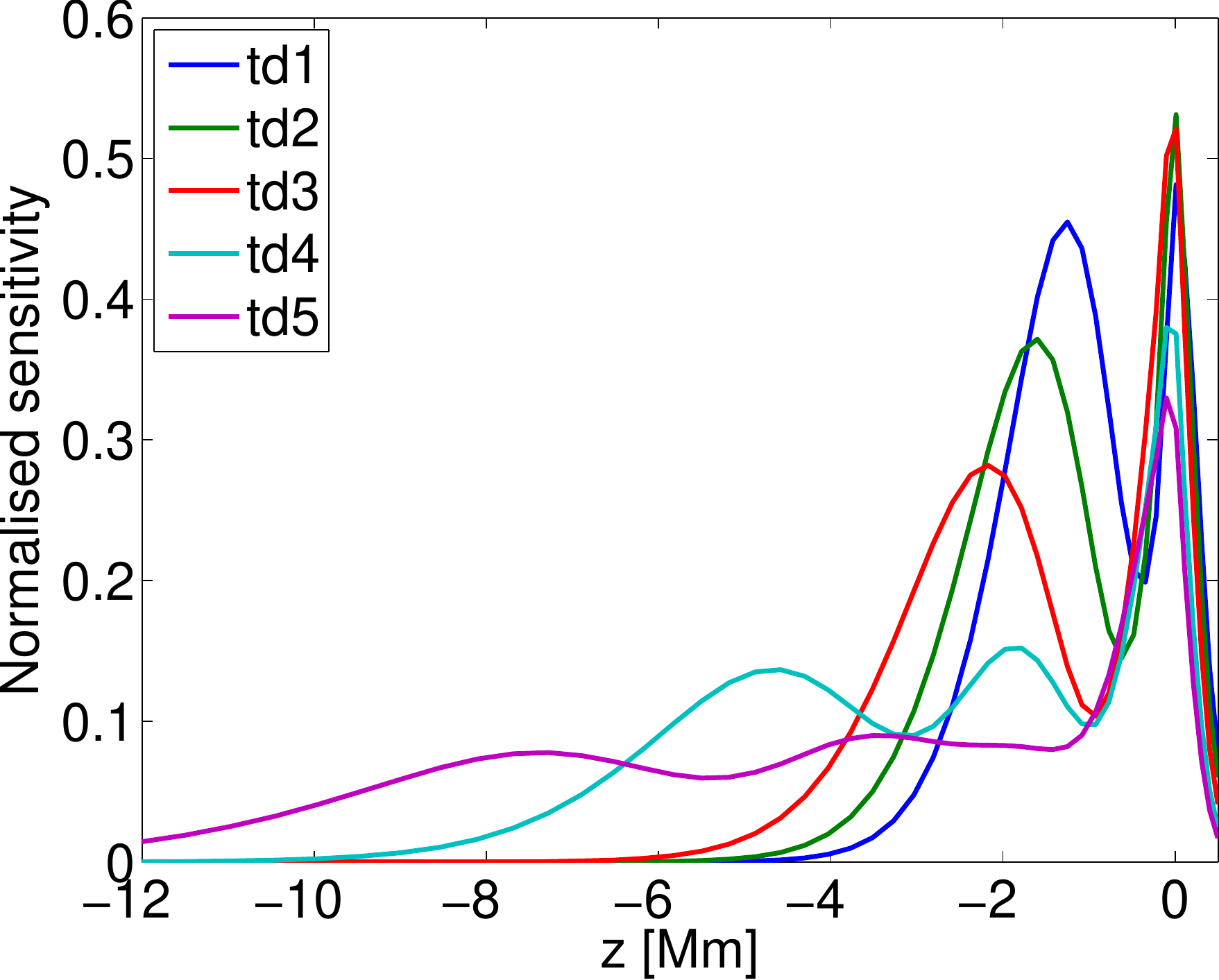}
\includegraphics[width=0.5\textwidth]{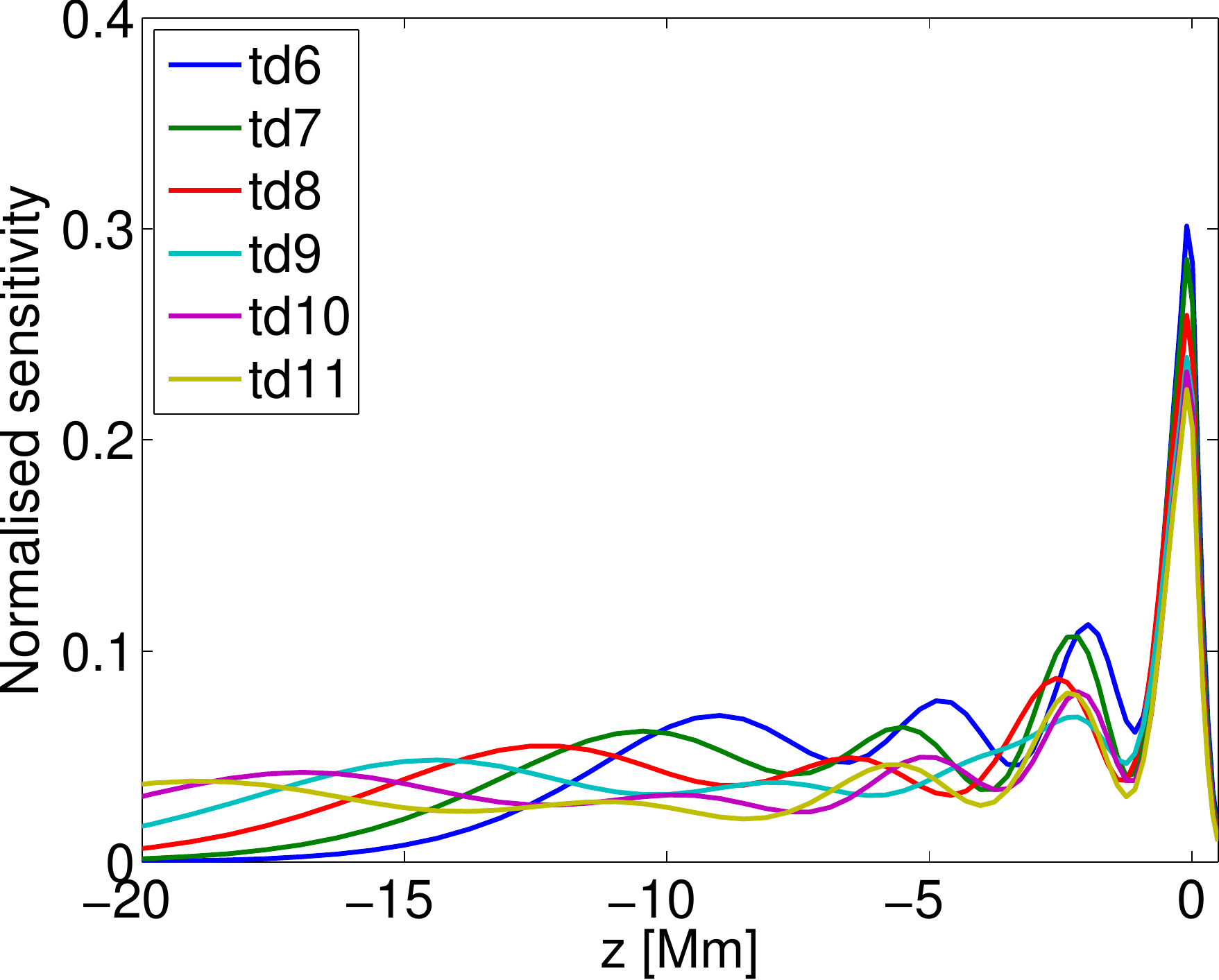}\\
\includegraphics[width=0.5\textwidth]{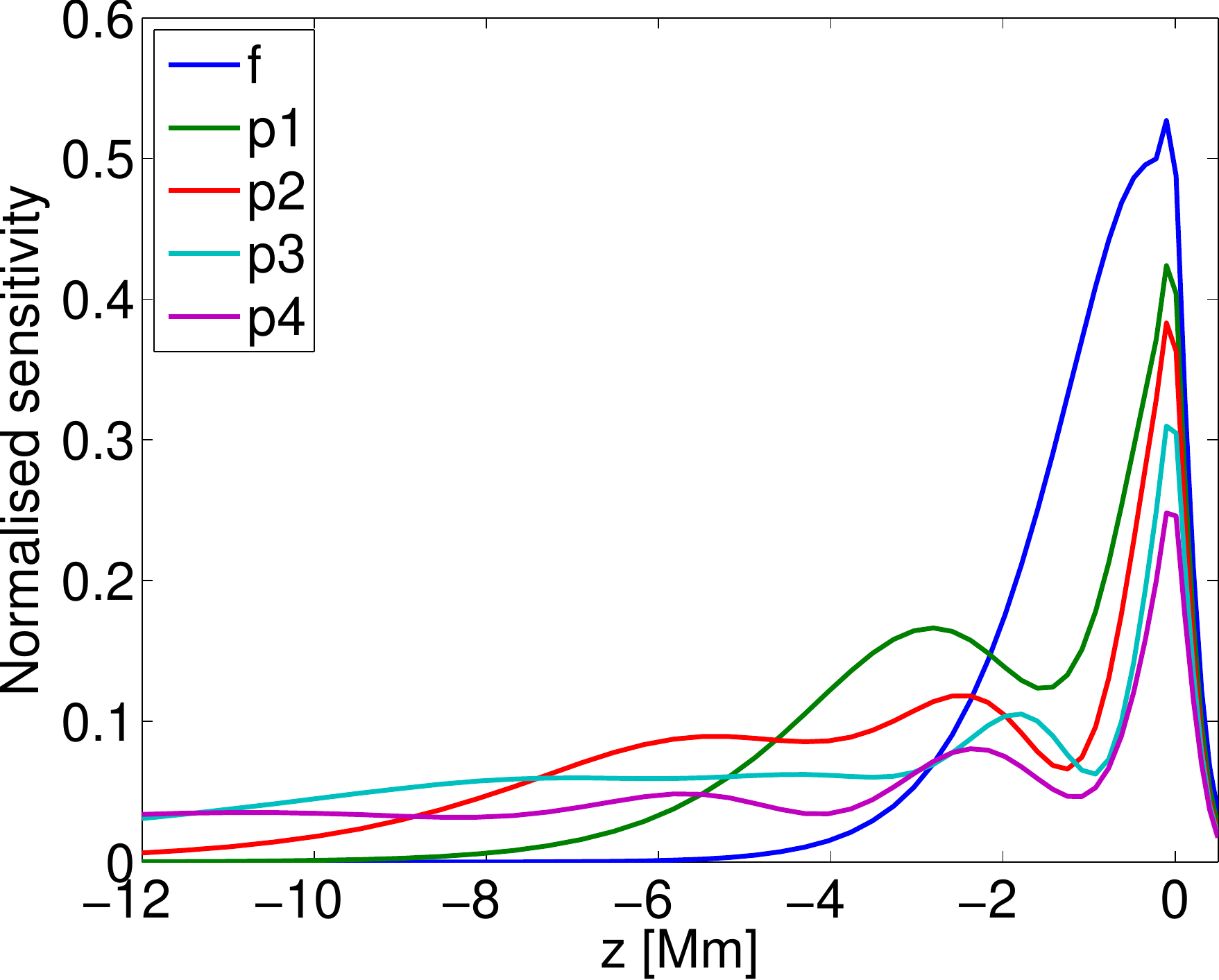}\\
\caption{The depth sensitivity of flow sensitivity kernels for spatio-temporal filters used in this study. One should note that all kernels are strongly sensitive towards the surface.}
\label{fig:1Dkernels}
\end{figure}

The inversion is performed using an implementation of the Multichannel SOLA \citep{fastOLA} in {\sc Matlab}, in details described by \cite{Svanda2011}. For the inversion, I selected a rather difficult test. The target depth was chosen almost randomly, 2.2~Mm, with full-widths-at-half-maximum (FWHM) of the Gaussian target function to be 2~Mm in vertical direction and 15~Mm in both horizontal directions. Looking at Fig.~\ref{fig:1Dkernels}, none of the filtered modes peaks at the chosen depth, some of the modes peak around it. Thus from the beginning, it is a difficult task for the inversion to produce meaningful results. 

Two inversions for horizontal components of the flow with the above described target function were executed, one employing all ridge-filtered sensitivity kernels ($f$ to $p_4$, together utilising 240 independent measurement geometries $a$) and another one with all phase-speed-filtered sensitivity kernels (TD1 to TD11, together employing 165 independent measurement geometries $a$). The targeted noise level in the inversion was 35~\mps{} and less for travel times averaged over one day, other requirements put on the inversion were to have spatially confined weights $w_a^\alpha$ and rather minimised crosstalk components of the averaging kernel ($\cK_\beta^\alpha$ for $\beta \ne \alpha$). 

\begin{figure}
\includegraphics[width=0.9\textwidth]{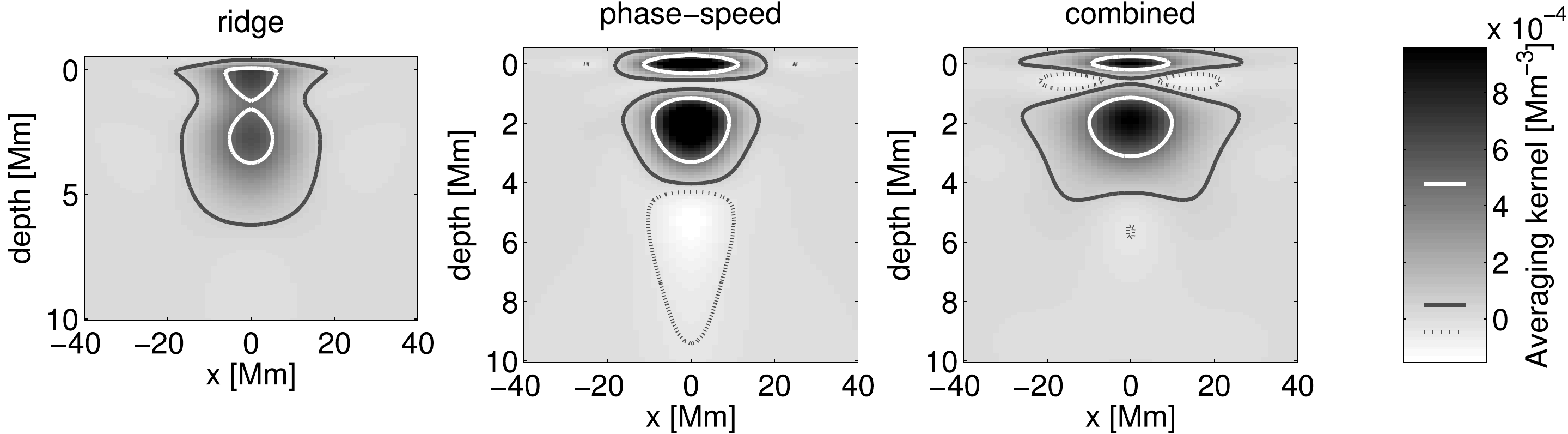}
\caption{Cuts along $y=0$ of averaging kernels $\cK_x^x$ for $v_x$ inversion at depth of 2.2~Mm using various filtering approaches. Over-plotted contours,
 which are also marked on the colour bar for reference, denote the following: half-maximum of the kernel (white), and $\pm 5$\% of the maximum value of the kernel (gray solid and gray dashed lines, respectively).}
\label{fig:test1}
\end{figure}

\begin{figure}
\includegraphics[width=0.5\textwidth]{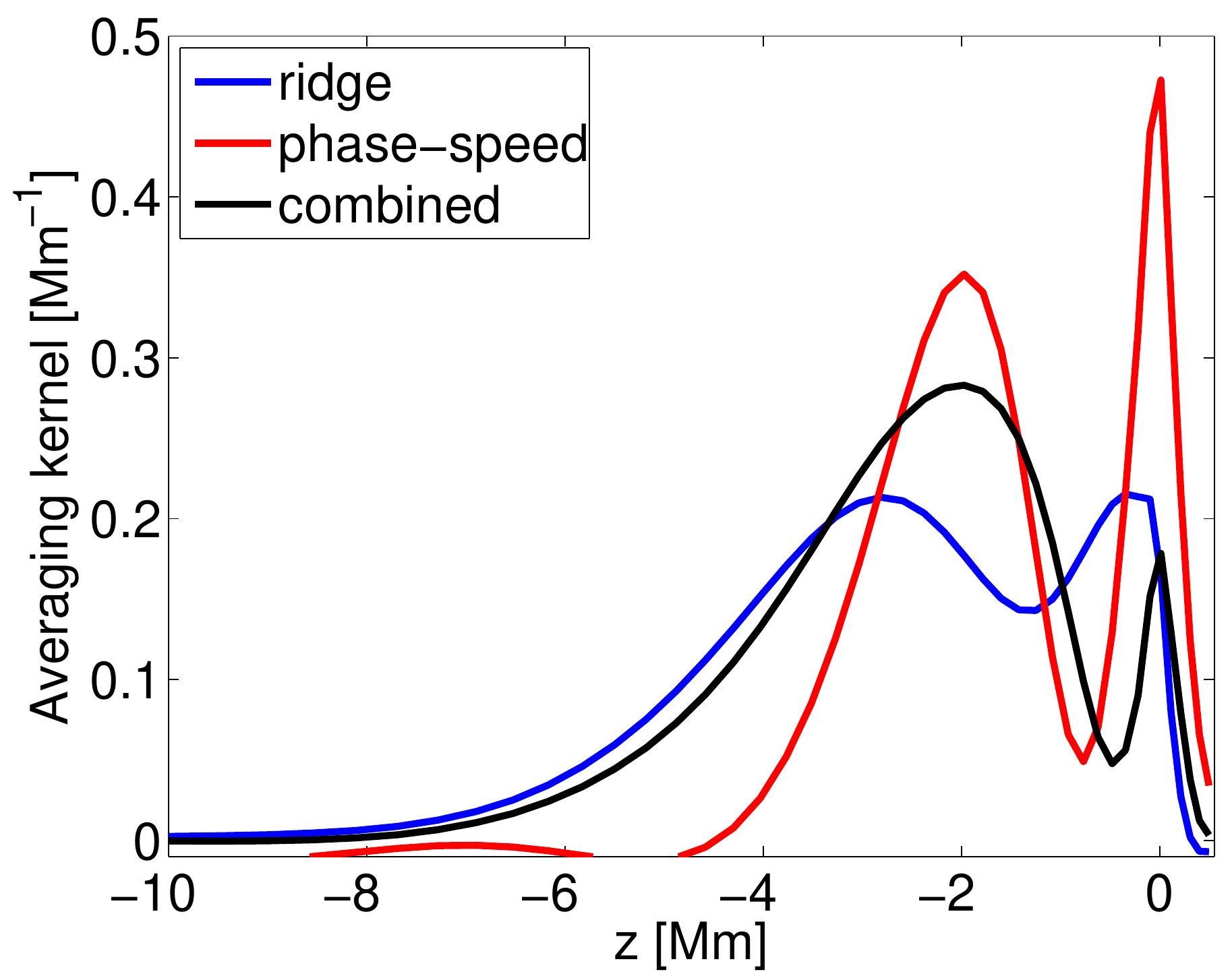}
\includegraphics[width=0.5\textwidth]{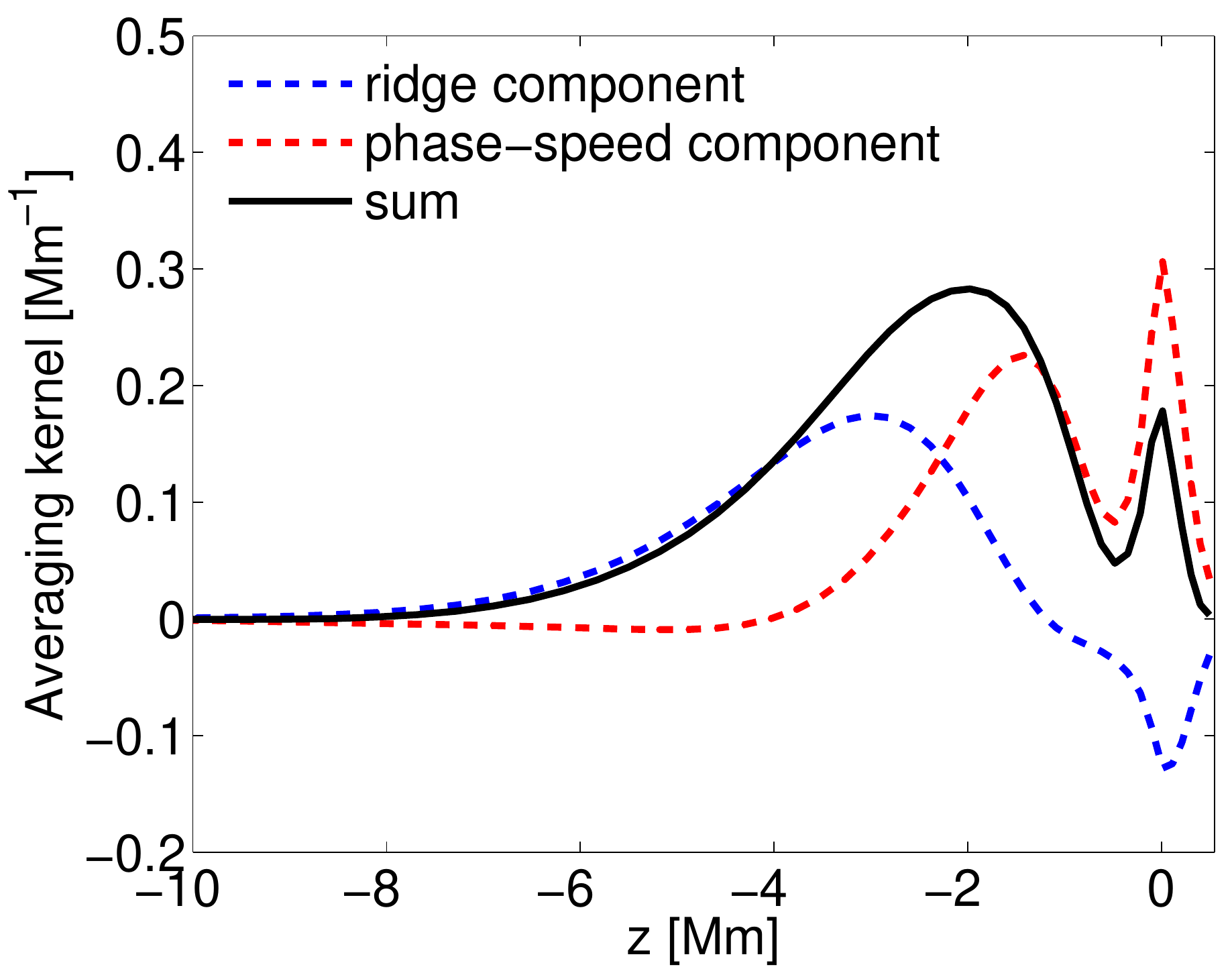}
\caption{Left: Horizontally averaged averaging kernels for $v_x$/$v_y$ inversion targeted at depth of 2.2~Mm using various $k$--$\omega$ filtering approaches. Right: Contributions of the ridge filters (blue) and phase-speed filters (red) to the averaging kernel of the combined inversion.}
\label{fig:test2}
\end{figure}

\begin{figure}
\includegraphics[width=0.9\textwidth]{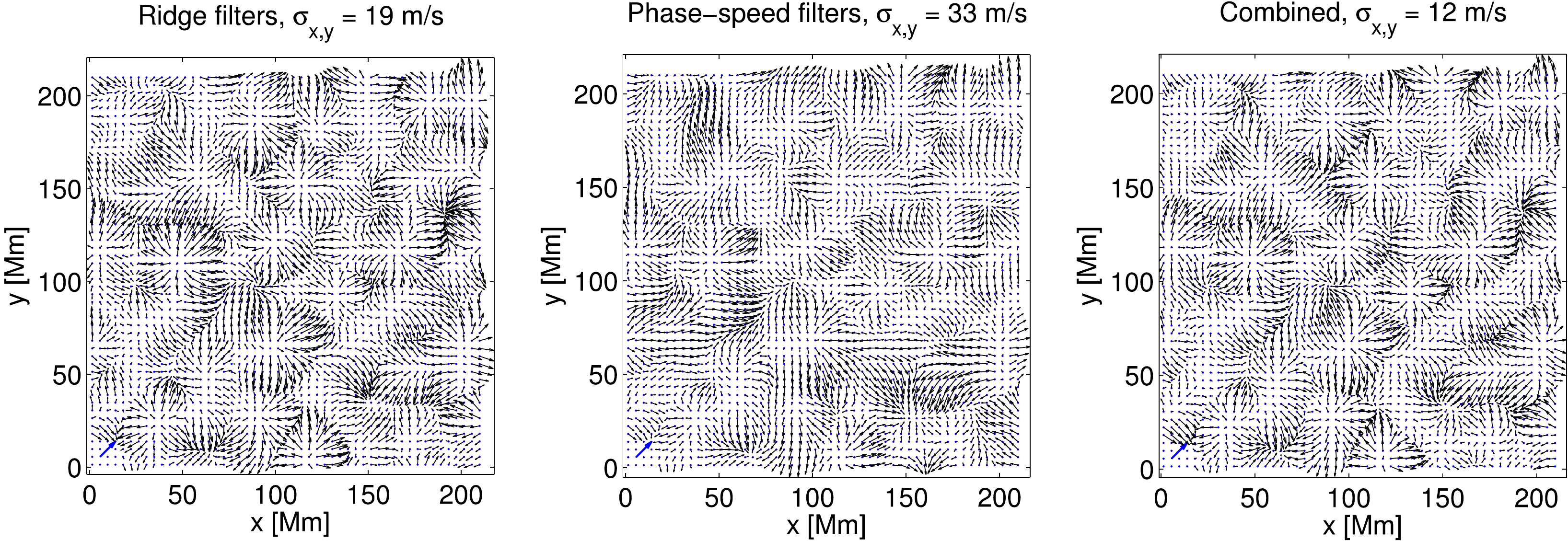}
\caption{Inferred horizontal velocity vector at depth of 2.2~Mm using various $k$--$\omega$ filtering approaches. The noise levels are given in the titles of panels. The reference arrow has length of 250~\mps.}
\label{fig:test3}
\end{figure}

Nevertheless, both $k$--$\omega$ filtering approaches should, in principle, provide us with similar estimates for horizontal flow velocities. That was confirmed when the inversions were validated by utilising the synthetic data using approach identical to \cite{Svanda2011}. The results of inversions applied to real observations are displayed in Figs.~\ref{fig:test1}--\ref{fig:test3}. The noise levels in the results are 19~\mps{} in the case of inversion combining the ridge-filtered measurements and 33~\mps{} in the second case. One should notice, e.g. on Figs.~\ref{fig:test1} and \ref{fig:test2} (left panel), that the inversion averaging kernels are not free of side-lobes, which always makes the proper interpretation of the results difficult.

Despite the side-lobes, averaging kernels for both inversions scan similar depths (although ridge-filtered inversion averages a little more signal from depth 3--6~Mm than the phase-speed-filtered inversion), so one should expect the resulting horizontal flow to be highly correlated. It is also evident, especially from looking at Fig.~\ref{fig:test2} left, that one should expect the scaling of inverted flow magnitudes. The phase-speed-filtered inversion depicts an increased sensitivity towards depths 0--1~Mm and 1.5--3~Mm compared to the ridge-filtered inversion. This increased sensitivity at the near-surface depths is compensated by a large negative side-lobe at depths 5--10~Mm, so that both averaging kernels are normalised by an explicit constraint incorporated into the inversion: $\int_\odot \id^2\bvec{r} \id z\, \cK_x^x = 1$. Assuming that the amplitude of the convection speed decreases sharply with depth, as numerical simulations \citep[e.g.][]{2008ASPC..383...43U} suggest, the deep negative side-lobe does not compensate for the increased sensitivity of the phase-speed-filtered inversion in the near-surface layers. The ratio of total integrals of averaging kernels over the near-surface depths should therefore provide us with an estimate of the expected scaling factor of magnitudes of inverted velocities resulting from both inversions. This ratio is 1.4 for the depth 0--3~Mm and 1.2 for 0--4~Mm, thus one should expect a magnitude scaling factor in this order. 

The inverted flow fields are displayed in the left and middle panels of Fig.~\ref{fig:test3}. Already a comparison by eye reveals that the inverted flows are highly correlated and the statistical analysis (see Table~\ref{tab:test}) largely confirms this visual impression. The scaling factor discussed in the previous paragraph evaluated by means of the least-squares fit to the flow estimates using error levels in both variables \citep{2004AJF..72...367Y} at all points in the field of view takes the value of 1.45 in the case of inversion for $v_x$ and 1.53 for the inversion for $v_y$, thus consistently with the predictions based on the analysis of averaging kernels discussed above. 

This simple test showed that the flow estimates inverted using different $k$--$\omega$ filtering approaches are very consistent. The experiment showed some of the difficulties of both filtering approaches (e.g. a large sharp surface side-lobe in case of the phase-speed filtering), possibly complicating the interpretation of the results. 

The test performed showed a substantial advantage of using ridge filters at the discussed depth. The averaging kernel is cleaner in case of the ridge filters and the noise level is simultaneously better, compared to similar inversion using the phase-speed filters. From other tests performed, this seems to be always the case within very shallow sub-surface layers. There are two effects possibly explaining this fact. The ridge-filtered inversion involves more independent measurements (240) than the phase-speed-filtered inversion (165). This provides more freedom for the inversion to find a better ballance between the terms in the cost function. On the other hand, it is evident from Fig.~\ref{fig:1Dkernels} that the overall depth sensitivity of the ridge-filtered kernels is more confined to the shallow sub-surface layers than the sensitivity of phase-speed-filtered kernels. When targeting the shallow layers using ridge filters, the sensitivity at larger depths naturally does not contribute much to the misfit term $||\bcK-\bcT||$ in the cost function and the algorithm may, in a shortcut, ``focus to find a better fit at the depths of interest''. In case of phase-speed filters the larger depths do contribute the misfit term and the cost function minimisation has consequences to both the quality of the averaging kernel at depths of interest and the level of random noise. Perhaps a differently constructed misfit term than using a simple norm might improve this issue. From the same reasons, the conclusions obtained using different inversion schemes, e.g. Multiplicative OLA \citep{1968GeoJ...16..169B} or Regularised Least Squares \citep{1982ACMTM...8...43P}, may be different.

\subsection{Inversion combining both filtering approaches}
\label{sect:combination}
Let's try to combine both $k$--$\omega$ filtering approaches in a one large inversion, thus involving 405 independent measurement geometries (index $a$)\footnote{Such computation is rather demanding. Inversion for one set of three trade-off parameters \citep[full details in][]{Svanda2011} balancing the misfit of the averaging kernel and the target function, the level of random noise, the level of cross-talk contribution, and the spatial localisation of the weights computed with 200$\times$200 wave vectors takes around 6 hours on a 2.6-GHz Opteron CPU. I usually cycle over 200 various combinations of trade-off parameters values. The total memory requirement for such inversion using the code implemented in {\sc Matlab} is in the order of 700~GB. Thanks to the multichannel approach \citep{fastOLA} to solve the inverse problem (under the assumption of homogeneous background, which is a reasonable assumption in the quiet Sun regions) the task is embarrassingly parallelisable in the wave-vector space, which also decreases the memory requirements of individual parallel jobs. Thus, using the {\sc Sunquake} cluster at Astronomical Institute of Academy of Sciences of the Czech Republic (32 CPU cores in total and 352 GB RAM in total) a typical inversion run for one target function takes around 60 hours.}. The results are plotted in Figs.~\ref{fig:test1}--\ref{fig:test3} for comparison with the previous two methods of $k$--$\omega$ filtering. The averaging kernel is better localised around the target depth (it does not have the very extended surface side-lobe as in the case of ridge filtering and does not have the extended deep negative side-lobe as in the case of phase-speed filtering, see Figs.~\ref{fig:test1} and \ref{fig:test2}). At the same time, the noise level in the results is lower (12~\mps{} compared to 19~\mps{} or 33~\mps{} respectively). Not surprisingly, the inverted flow is highly correlated with the flow estimates from both ridge-filtered and phase-speed-filtered flow estimates (see Table~\ref{tab:test}). In the inversion, most of the modes contribute the results, except for the $p_4$ ridge, which has significantly lower contribution, and phase-speed filters TD8 to TD11, which have more than an order of magnitude lesser contribution. 

The correlation coefficients suggests that the inversion combining both filtering approaches is largely dominated by the ridge filters. Although a substantial contribution to the main lobe of the averaging kernel comes from the ridge filters, phase-speed filters contribute comparably (Fig.~\ref{fig:test2} right). The lower correlation with the solely phase-speed-filtered inversion is thus a consequency of a larger random-noise level, which is almost twice larger in case of the phase-speed-filtered inversion than in case of the ridge-filtered inversion. The substantially lower level of the random noise in the combined inversion might be surprising as both sets of measurements overlap over the $k$--$\omega$ diagram and thus one would not expect both measurements to be largely independent to help the inversion to constrain the noise level. Counter-intuitively to the expectations, the covariance matrices of the travel-time noise of the two measurement sets have values close to zero (yet, not exactly zero) indicating that both sets of measurements indeed are almost independent.

Obviously, the combination of more independent travel-time measurement geometries is an improvement to what was seen before. The power of this approach will become clear in the following application.

\section{Application: Tomography of sub-surface convection}
The ultimate goal of helioseismic inversions for flows is to study the structure of the flows deeper in the convection zone. However, as pointed out in Section~1, it has been difficult to obtain sensible flow measurements on supergranular scales from depths deeper than around 4~Mm. This is not deep enough to, for instance, answer the questions regarding the vertical structure of supergranulation cells. Let's hope for better when combining more independent travel-time measurements. As was shown in the previous Section, this approach seems feasible.

\subsection{Choice of optimal inversions}
\begin{figure}
\includegraphics[width=0.5\textwidth]{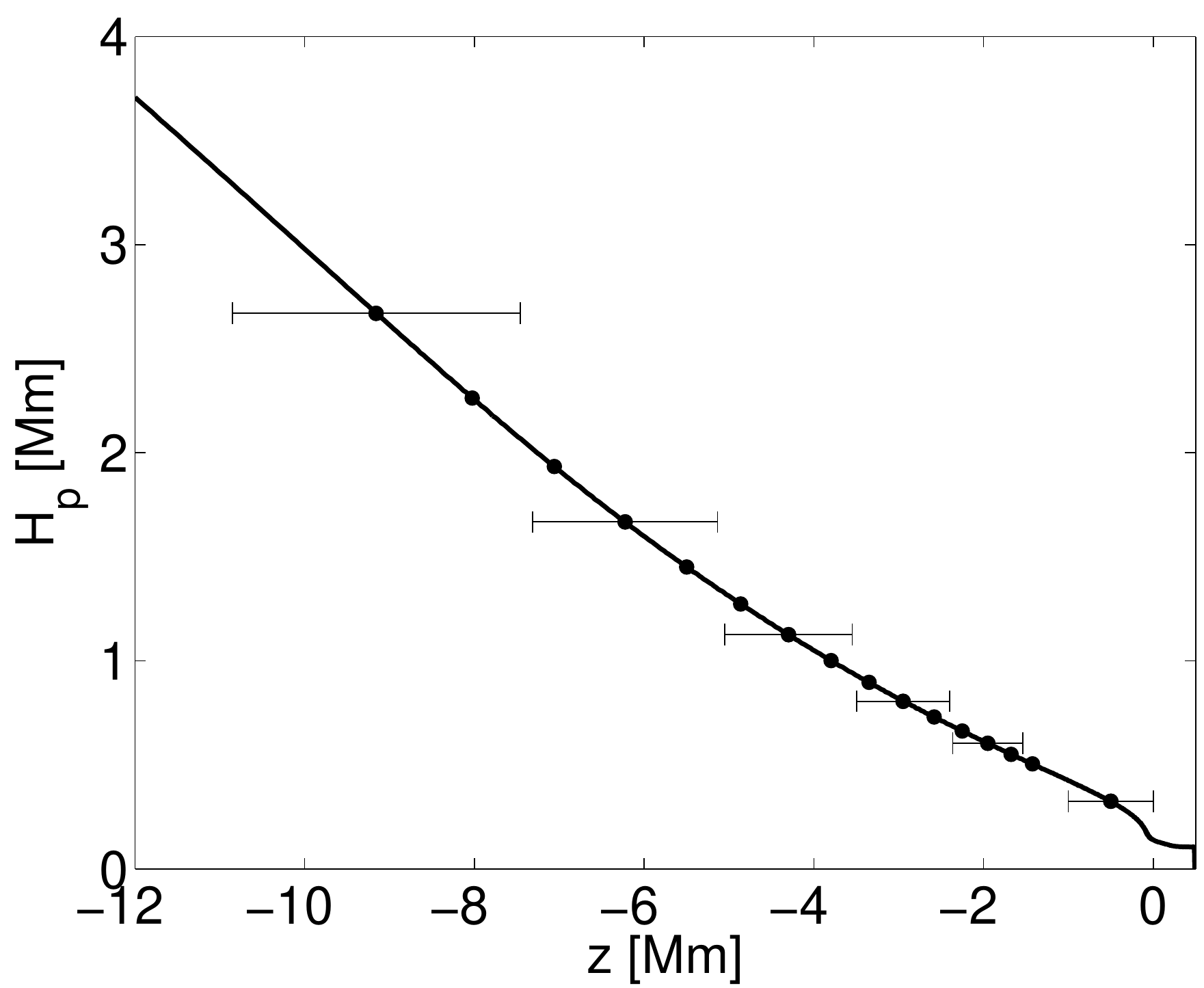}
\caption{Pressure scale height as a function of vertical coordinate. The separation of neighboring marked points on the curve is one half of the scale-height at the shallower point of the two. The range bars describe the parameters of the inversion target functions, indicating the location of the peak and FWHM of the Gaussian.}
\label{fig:targets}
\end{figure}

I ran a set of inversions for horizontal flow at six depths, target functions for whose were constructed on the physical basis. First, as a reference, I use the $f$-mode only inversion, which was validated against direct surface measurements \citep{Svanda2013} to represent the estimate of the near-surface horizontal flow. Then I proceed iteratively in depth, starting at depth of 1~Mm and move deeper always by one half of pressure scale height. Each third step in this iterative sequence represents one target depth for the inversion, the 1.5-multiple of the pressure scale height at this depth is then used as FWHM of the Gaussian target function (see Fig.~\ref{fig:targets}). The starting depth of the sequence was chosen because according to \cite{2006ApJ...640..516C} it makes no sense to scan for features extending less than 1~Mm in vertical direction. I do not go deeper than 10~Mm. Thus, in the end, there are six depths, target functions for which are sort of independent (they do not overlap withing their respective FWHMs) -- surface (thereafter denoted as 0~Mm), $1.9$, $2.9$, $4.3$, $6.2$, and $9.2$~Mm. All these inversions were validated using synthetic data following the approach of \cite{Svanda2011}.

A curious reader familiar with all the detail of the cost function of the inverse problem \citep{Svanda2011} may ask a perfectly valid question, how did I select the proper combination of three trade-off parameters (in Section~\ref{sect:combination} I already wrote that a typical inversion run contains 200 of combinations of those trade-off parameters) for the inversions presented further. A correct answer to this question is that it is based on a trial-and-error approach. 

I use a three-step manual down-selection. The first eliminating step consists of bounding the level of random noise in the results, predicted by the inversion. Here I used the error level to be around 40~\mps{} for all inversions assuming averaging time of 24~hours. It is my requirement to have all inversions with comparable noise levels This limit was fulfilled for all depths, except for 0~Mm (noise level of 18~\mps{} for the given averaging time), where I strictly used the validated surface inversion \citep{Svanda2013}, and the depth of 9.2~Mm, where I had to relax the noise constraint to value of around 50~\mps, otherwise the averaging kernel had too many side-lobes. The first step eliminates most of 200 trade-off parameter combinations and retains usually around twenty solutions. 

The second step involves looking at the inversion weights $w_a^\alpha$. The weights must be strongly localised around the central point in the spatial domain and the experience says that they should have maximal values of the order of $10^{-5}$~km\,Mm$^{-2}$\,s$^{-2}$. Values one order larger, which are also often seen, provide usually inversion which is not robust, as the weights oscillate. Should any of the set of weights be rather extended in the spatial domain, the solution is rather periodic in spatial domain and strongly limits the useful field of view. After this step, usually less than five acceptable solutions remain.

The last criterion I evaluate is the look at the resulting averaging kernels. From my experience, it is vastly more important to have an averaging kernel, which is without side-lobes or have at least side-lobes as minimised as possible (this is imperative especially for negative lobes), than an averaging kernel that is close to the inversion target function. Thus a simple selection based on the misfit $||\bcK-\bcT||$ does not necessarily provide the best solution. In this step I also search for an averaging kernel with cross-talk components $\cK^\alpha_\beta$ for $\beta \ne \alpha$ minimised. 

\begin{figure}
\includegraphics[width=0.5\textwidth]{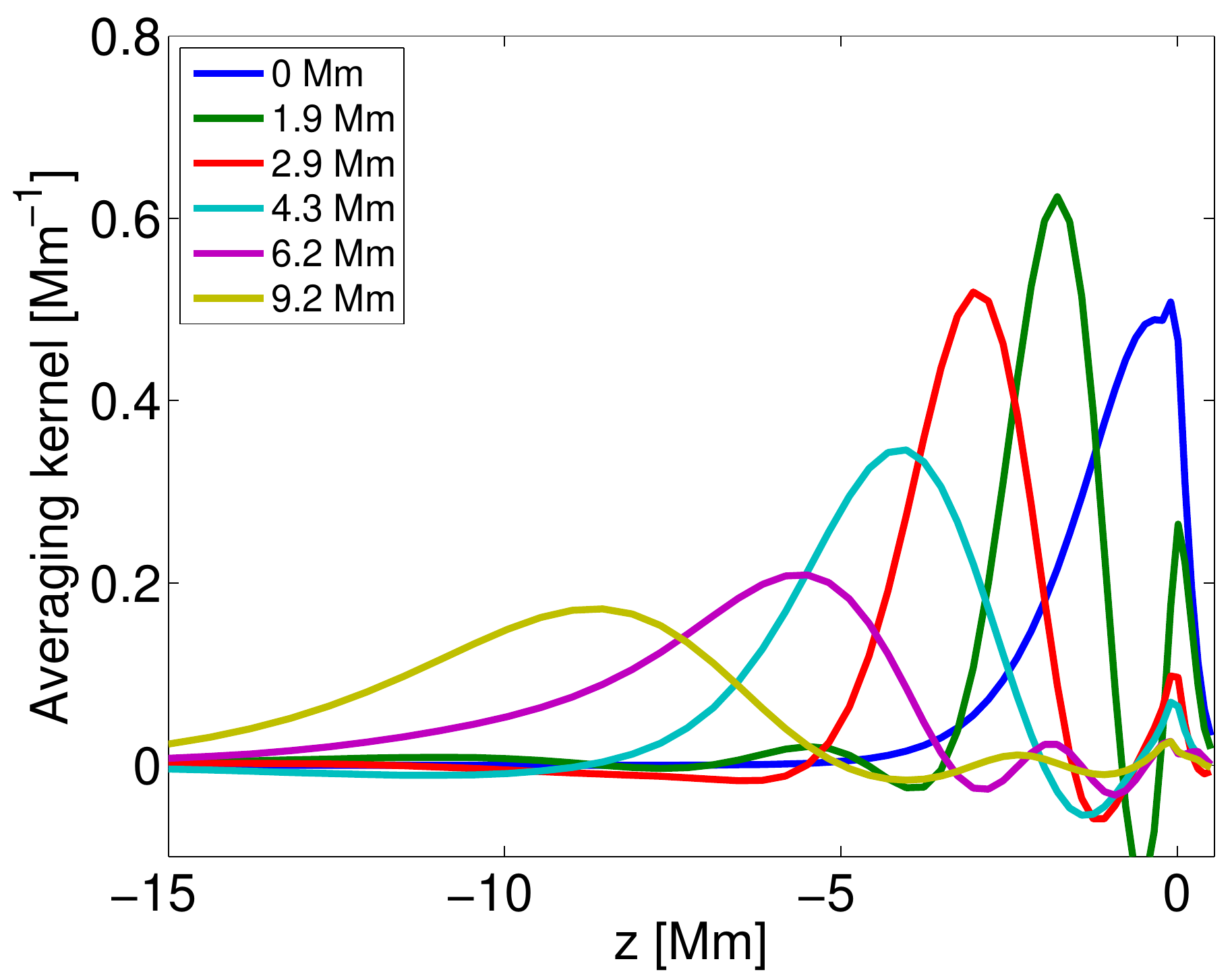}
\caption{Horizontally averaged averaging kernels of inversions for horizontal flows components at six depths in the solar convection zone. }
\label{fig:tomography_kernels}
\end{figure}

The parameters of the selected inversions are summarised in Table~\ref{tab:tomography_parameters} and the resulting averaging kernels displayed in Fig.~\ref{fig:tomography_kernels}. The horizontal FWHM of the target functions is always 15~Mm and is well reproduced by the averaging kernels. One should notice that the averaging kernel for depth of 1.9~Mm has large positive and negative near-surface side-lobes, making the interpretation of the results somewhat difficult. 

\begin{figure}
\includegraphics[width=0.5\textwidth]{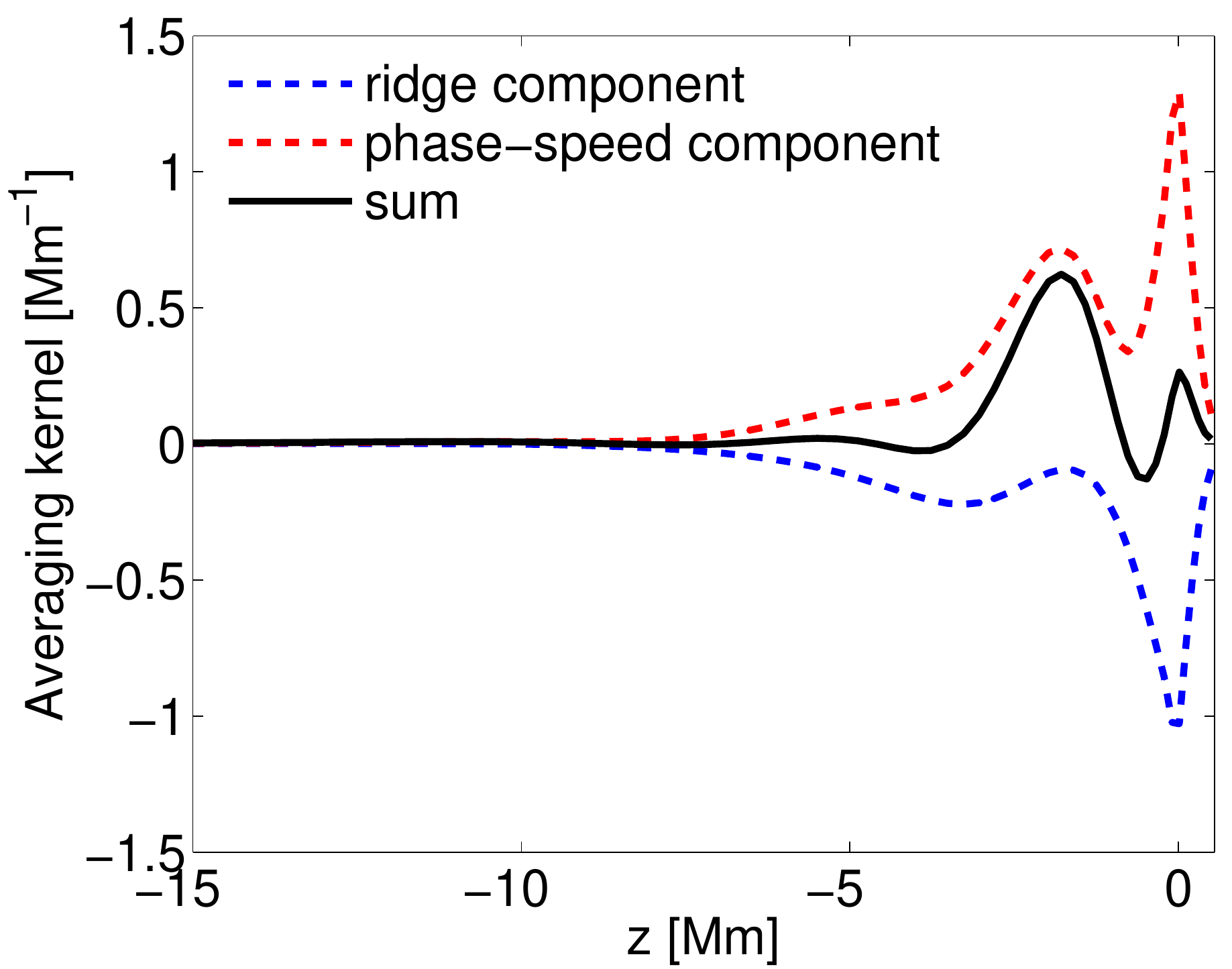}
\includegraphics[width=0.5\textwidth]{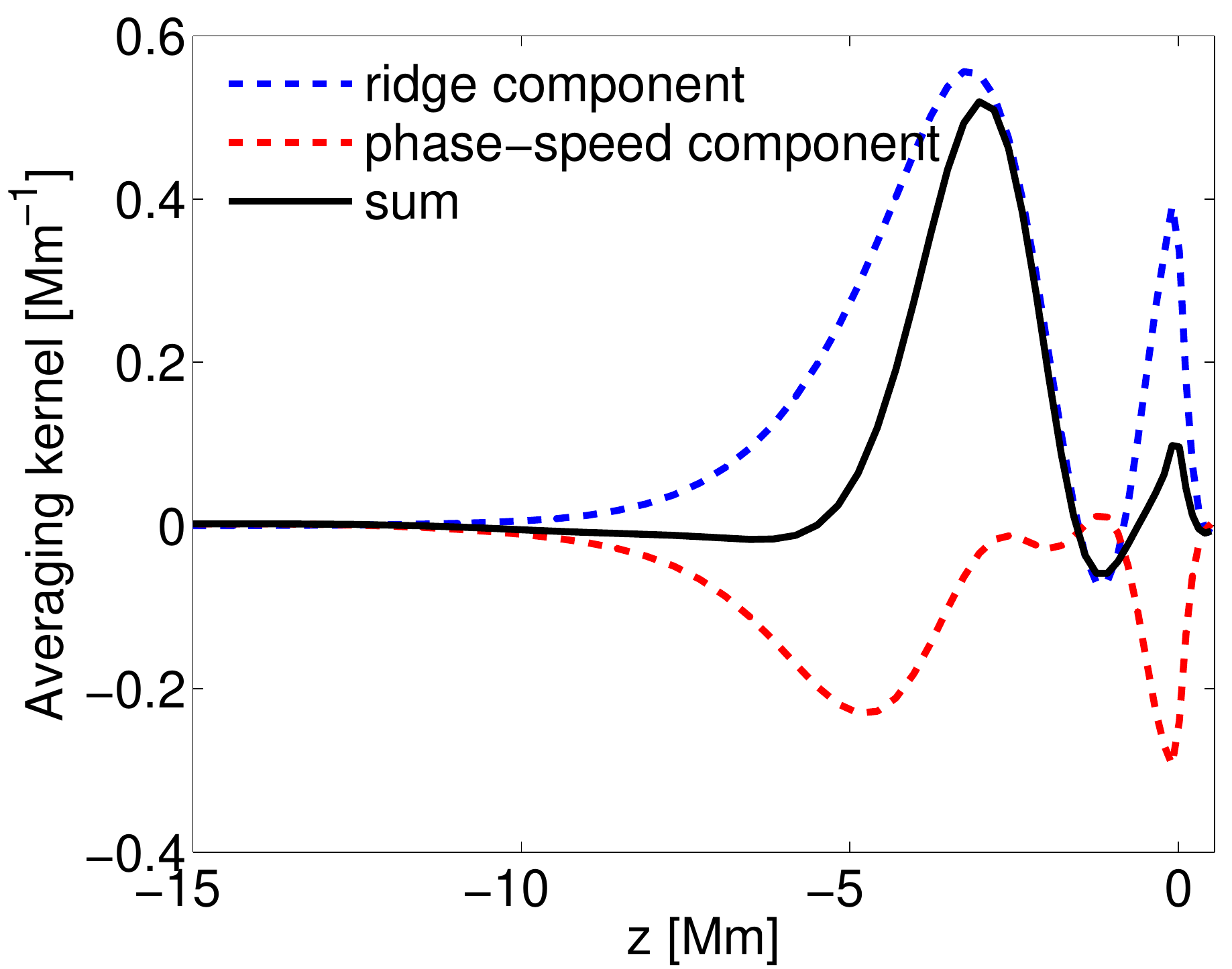}\\
\includegraphics[width=0.5\textwidth]{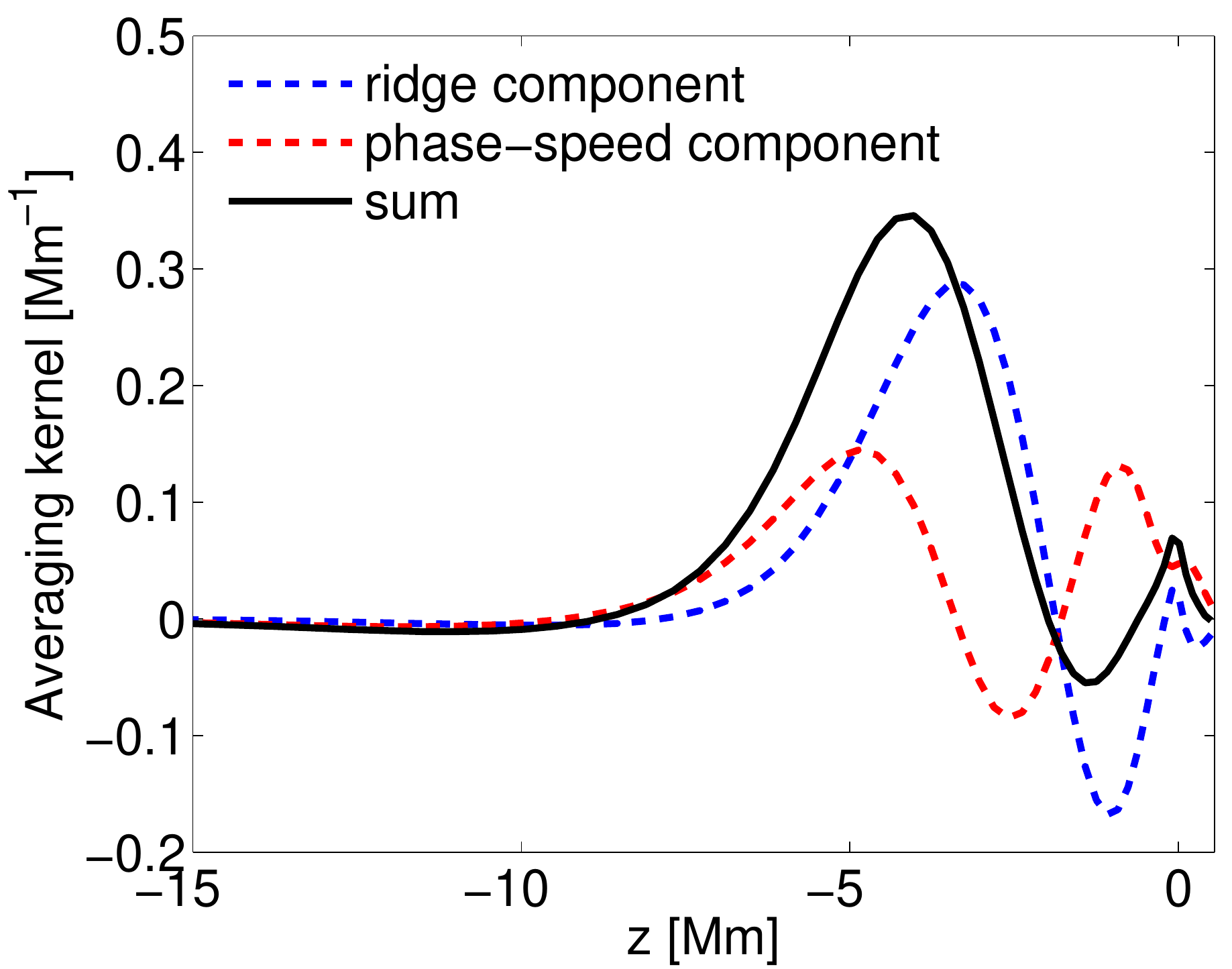}
\includegraphics[width=0.5\textwidth]{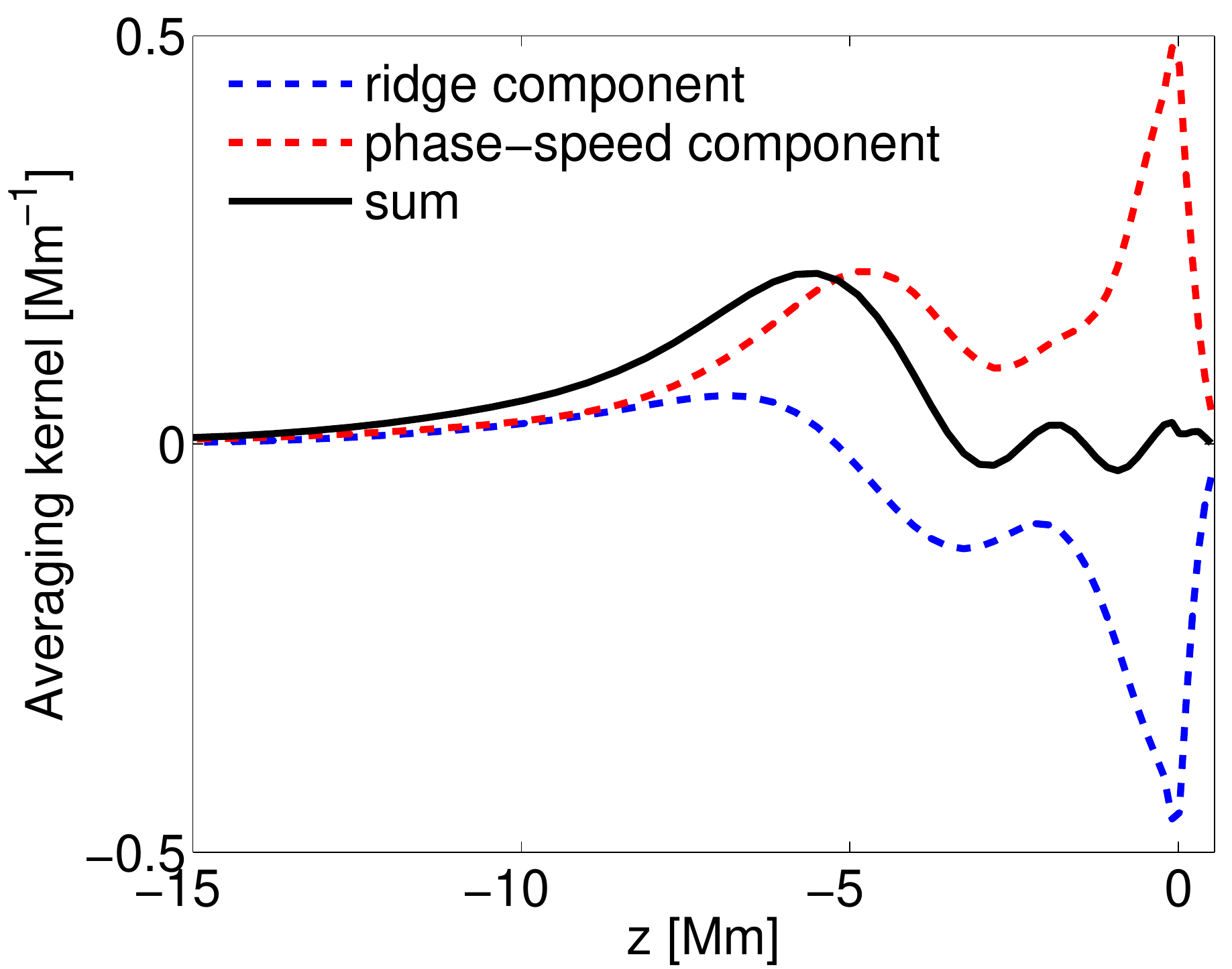}\\
\includegraphics[width=0.5\textwidth]{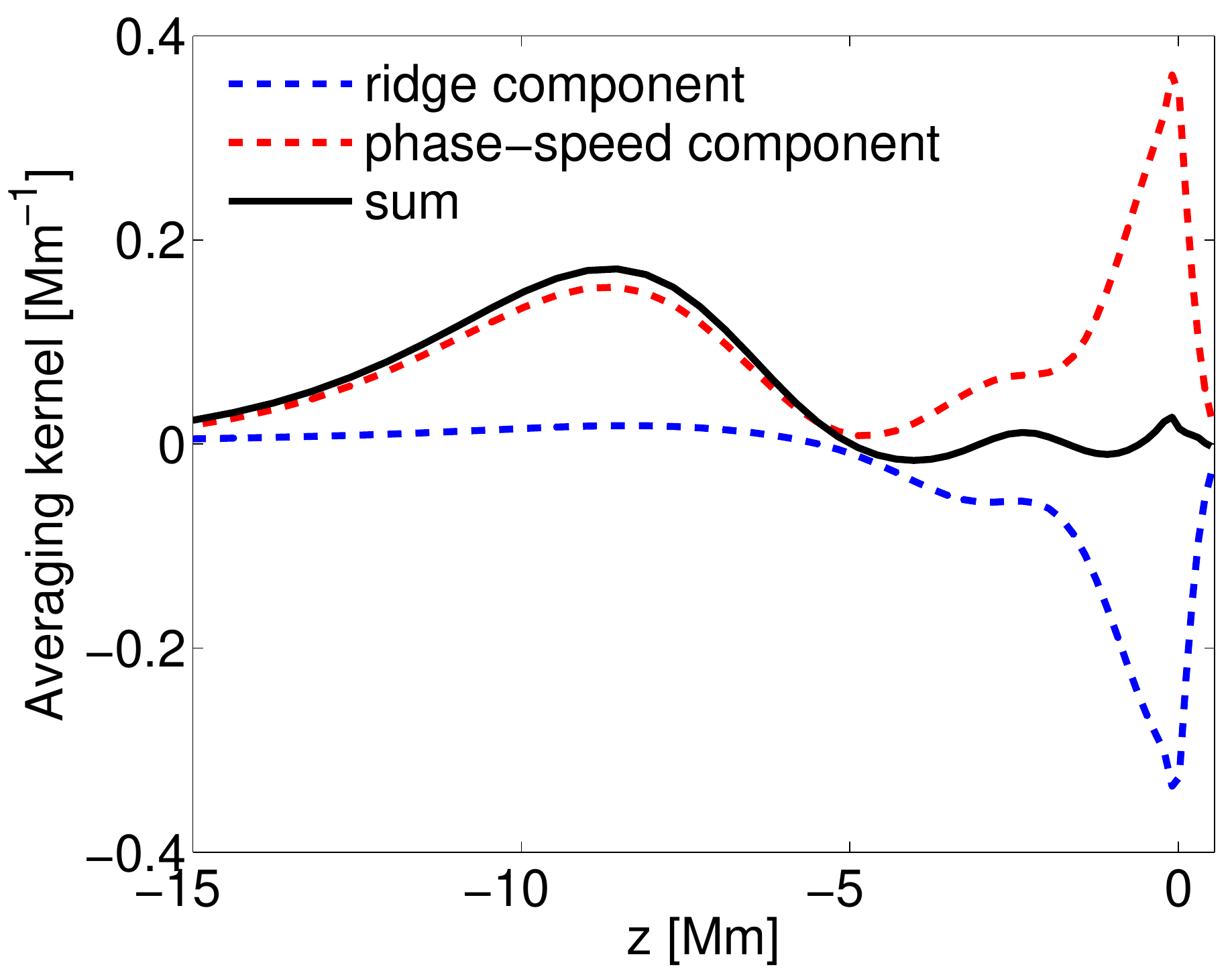}
\caption{Contributions of the ridge filters (blue) and phase-speed filters (red) to the averaging kernel of the combined inversions at depths $1.9$, $2.9$, $4.3$, $6.2$, and $9.2$ respectively. One sees an increasing importance of the phase-speed filters with depth.}
\label{fig:composition}
\end{figure}

With depth, the importance of the phase-speed filters to construct the main lobe of the averaging kernels increases rapidly (see Fig.~\ref{fig:composition}). 
The ridge filters on the other hand help to eliminate the near-surface sidelobe. In a shortcut, for inversions deeper than say 5~Mm, the phase-speed filters form the main lobe of the averaging kernel, while ridge filters help to constraint the shallow layers. For inversions shallower than 5~Mm no such conclusion can be drawn.

\subsection{Properties of inverted flows}
\begin{figure}
\includegraphics[width=0.99\textwidth]{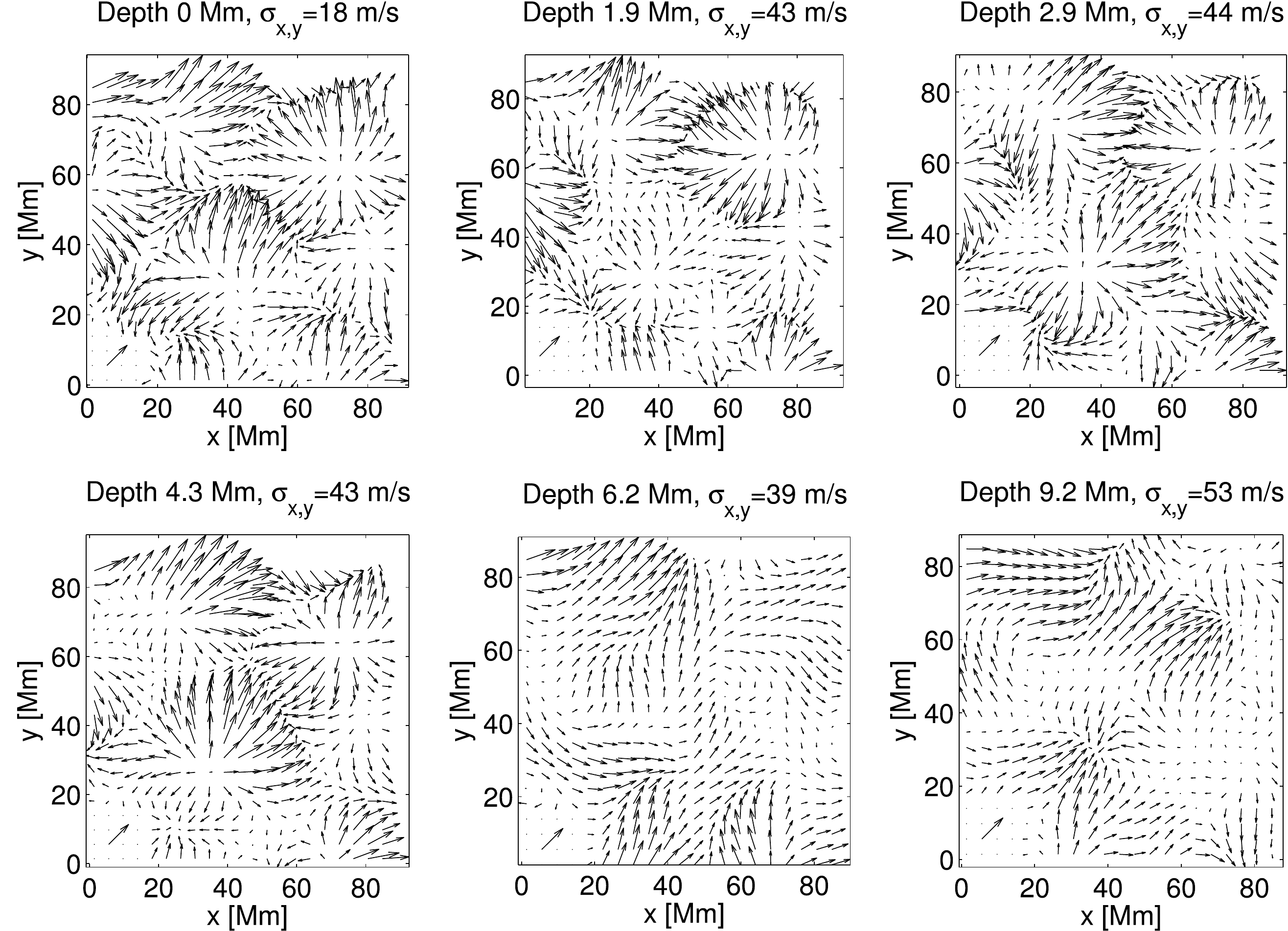}
\caption{An example selection of the measured flow maps at six different depths in a randomly selected section of the field of view. Scaling arrow indicates the flow of 200~\mps.}
\label{fig:tomography_example}
\end{figure}

An example of the inverted flows for one of nine investigated days is displayed in Fig.~\ref{fig:tomography_example}. It is evident that in the near-surface layers, the flow field is dominated by the plasma flows within supergranules, having a typical size of some 30~Mm. The visibility of supergranules drops rapidly in the flow maps inverted at depths of 6.2 and 9.2~Mm. This is a common picture for all nine investigated days. Visual inspection of flow maps in these depths actually does not reveal any clear cellular pattern as the one clearly visible in shallower depths. There are, however, divergent and convergent centres even at large depths, which do not seem to be separated by some typical distance. Given the predicted error and root-mean-square velocities at these levels (see Table~\ref{tab:tomography_parameters}), the inverted velocities have signal-to-noise ratio larger than one, so the poor visibility of supergranules deeper than 6~Mm does not seem to be caused by signal drowned in random noise. 

The mutual correlation coefficients for inverted flows at all investigated depths calculated for all nine sets of flow maps are given in Table~\ref{tab:tomography_correlation}. A high correlation between depths of 0~Mm, 1.9~Mm, 2.9~Mm, and 4.3~Mm is evident, thus one can conclude that supergranules are coherent structures within these depths. A slight anti-correlation is seen at larger depths, in agreements with some previous studies \citep[e.g.][]{1998ESASP.418..581D,2003ESASP.517..417Z,2009NewA...14..429S,2011NewA...16....1Z}. The negative correlation was usually interpreted as the detection of the flow reversal within the supergranular cells. 

\begin{figure}
\includegraphics[width=0.5\textwidth]{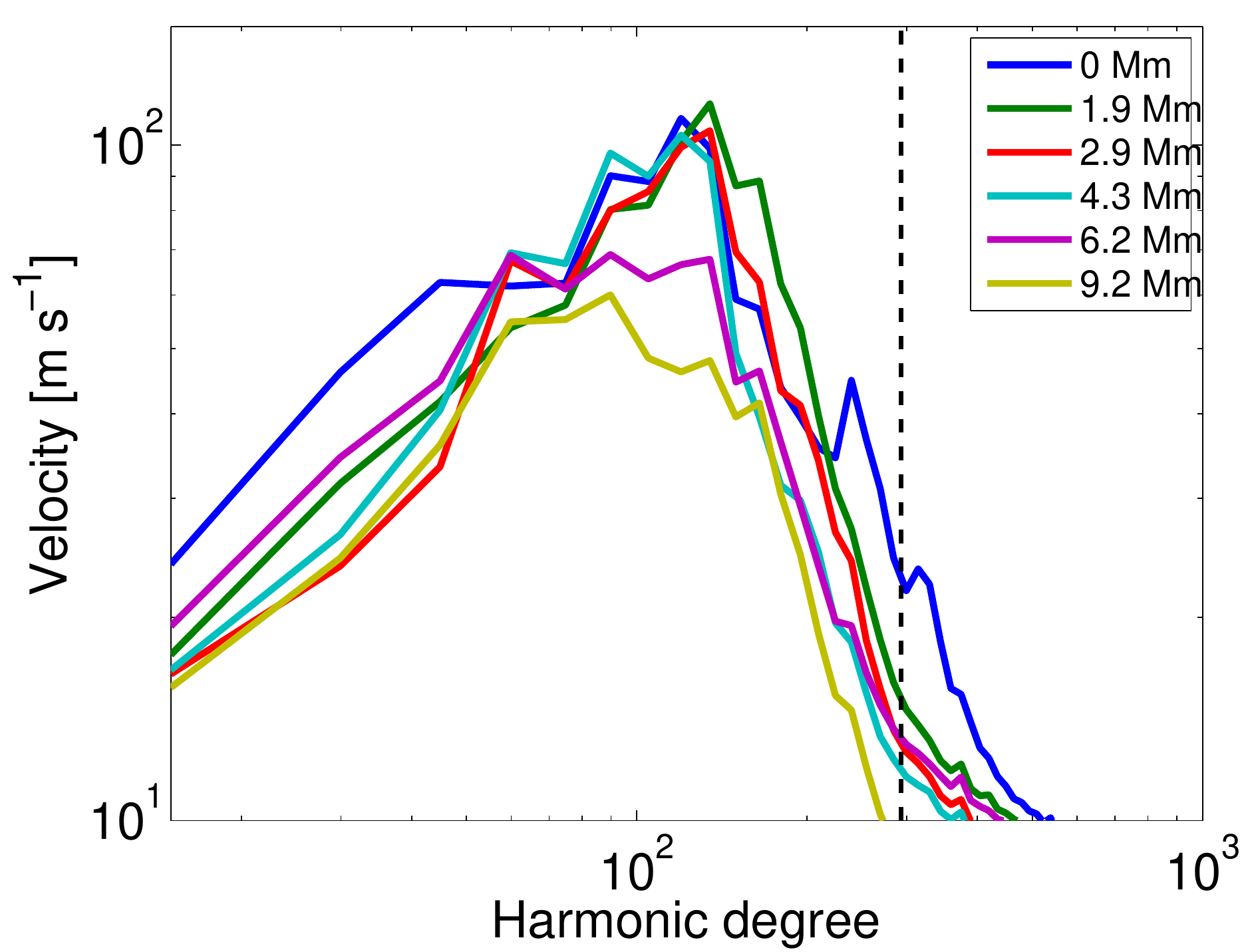}
\caption{Velocity spectra as a function of harmonic degree derived from horizontal flows measured at six analysed depths. The spectra were averaged over nine investigated datacubes each averaged over 24 hours. The dashed vertical line indicates the spatial cut-off introduced by the horizontal part of the averaging kernel, thus scales approaching this limit are effectively filtered out.}
\label{fig:tomography_spectra}
\end{figure}

Let's look at it at a little more details. When computing the correlation coefficient and especially when interpreting its large positive or negative values, one assumes that the studied structures remain at the same place and have a similar size in all correlated maps. Even a skewness of the structure in the depth domain will cause the correlation coefficient to drop, as will also the change in the size of the studied structures. To assess the latter issue, I constructed for all investigated depths the velocity spectrum $V(k)$, given by
\begin{equation}
V(k)=\sqrt{k P(k)},
\end{equation}
where $P(k)$ is the traditional power spectrum in absolute units \citep[for recipe I refer to][]{2010AA...512A...4R}. As discussed by \cite{2009LRSP....6....2N}, the velocity spectrum $V(k)$ is a good measure of the velocity amplitude at various scales. The velocity spectra, averaged over nine investigated 24-hours datacubes, are plotted in Fig.~\ref{fig:tomography_spectra}.

The velocity spectra obtained at different depths indicate the shift of typical scale of velocity structure from spherical harmonic degree $l$ of 120 at the surface and depths down to 4.3~Mm to larger scales ($l \sim 80$) at the depth of 9.2~Mm. Let's remind again that the horizontal extent of the averaging kernel is the same (FWHM of 15~Mm) for inversions for all discussed depths, thus this should not have an effect on scales captured by the inversion. The negative correlation of deeper depths with the surface (reported also by other authors) may thus be spurious and only a consequence of loss of the coherence of supergranular cells with depth, together with action of random noise, which plays not negligible role at deeper depths. From the velocity spectrum, one sees also a hint on a decrease of the typical velocity magnitude with depth. At this stage, I remind that only 9 velocity spectrum realisations were averaged, which is not enough to draw any solid conclusion. This issue is one of the directions that will be taken in the future research.

\section{Final remarks}
I demonstrated that the two $k$--$\omega$ filtering approaches traditionally used in time--distance helioseismology provide within limitations consistent estimates of the near-surface flow field. The combination of these two filtering approaches in a single inversion provides an improvement to performance of inverse modelling of subsurface flows. It allowed to obtain tomographic images of horizontal flow on supergranular scale averaged over 24 hours at depths down to $\sim10$~Mm with signal-to-noise ratio greater than one. 

The preliminary analysis using only 9 days of observations without magnetic activity on the surface suggests a systematic shift of dominant scale of subsurface convection towards larger scales with increasing depth. We plan to investigate this phenomenon in a greater detail in the future by utilising much larger sets of observations.

\acknowledgements This work was supported by the Czech Science Foundation (grant P209/12/P568). All computations were performed using the {\sc Sunquake} compute cluster at Astronomical Institute of Academy of Sciences in Ond\v{r}ejov, the tracked and mapped datacubes were obtained from data processing pipelines at Max-Planck-Institut f\"ur Sonnensystemforschung (MPS), Katlengurg-Lindau, Germany, which is funded by the German Aerospace Center (DLR). The solar measurements were kindly provided by the HMI consortium. The HMI project is supported by NASA contract NAS5-02139. The point-to-point travel-time sensitivity kernels were obtained using the code of Aaron Birch deployed at the MPS. This work was carried out in collaboration with project SFB 963 ``Astrophysical  flow instabilities and turbulence'' (Project A1). Tato pr\'ace vznikla s podporou na dlouhodob\'y koncep\v{c}n\'\i{} rozvoj v\'yzkumn\'e organizace (RVO:67985815) a v\'yzkumn\'eho z\'am\v{e}ru MSM0021620860. Last but not least I thank the anonymous referee for valuable comments that helped to improve the message of this paper.


\clearpage
\begin{table}
\begin{tabular}{llll}
\hline
\hline
Filter & $\Delta$ [Mm] & $v_{\rm ph}$ [k\mps] & $\delta v_{\rm ph}$ [k\mps]\\
\hline
TD1 & 3.7, 4.95, 6.20, 7.45, 8.7 &  12.77 & 2.63\\
TD2 & 6.2, 7.45, 8.70, 9.95, 11.2 & 14.87 & 2.63\\
TD3 & 8.7, 10.15, 11.60, 13.05, 14.5 & 17.49 & 2.63\\
TD4 & 14.5, 15.72, 16.95, 18.17, 19.4  & 25.82 & 3.86\\
TD5 & 19.4, 21.87, 24.35, 26.82, 29.3 & 35.46 & 5.25\\
TD6 & 26.0, 28.27, 30.55, 32.82, 35.1 & 39.71 & 3.05\\
TD7 & 31.8, 34.27, 36.75, 39.22, 41.7 & 43.29 & 3.15\\
TD8 & 38.4, 40.67, 42.95, 45.22, 47.5 & 47.67 & 3.57\\
TD9 & 44.2, 46.67, 49.15, 51.62, 54.1 & 52.26 & 4.46\\
TD10 & 50.8, 53.07, 55.35, 57.62, 59.9 & 57.16 & 3.78\\
TD11 & 56.6, 59.12, 61.65, 64.18, 66.7 & 61.13 & 3.41\\
\hline
\end{tabular}
\caption{Annulus sizes, phase speed, and width of the filter for used eleven standard time--distance filters. Taken from Table 1 of \cite{2006ApJ...640..516C}.}
\label{tab:phs_filters}
\end{table}

\begin{table}
\begin{tabular}{r|ccc}
\hline
\hline
& ridge & phase-speed & combined\\
\hline
ridge & 1 & 0.79 & 0.93\\
phase-speed & 0.75 & 1 & 0.76\\
combined & 0.93 & 0.75 & 1 \\
\hline
\end{tabular}
\caption{Values of Pearson's correlation coefficient for estimates of the horizontal flow components using various $k$--$\omega$ filtering approaches. Above the diagonal of the table the correlation coefficients for $v_x$ are given, below diagonal then the correlation coefficients for $v_y$ estimates. }
\label{tab:test}
\end{table}

\begin{table}
\begin{tabular}{r|lllllll}
Depth&$\sigma_{x,y}$ & RMS($v_x$) & RMS($v_y$)& Note \\
& [\mps] & [\mps] & [\mps] & \\
\hline
\hline
0 Mm & 18 & 113 & 109 & $f$-mode only \\
1.9 Mm & 43 & 99 & 102  \\
2.9 Mm & 44 & 98 & 95  \\
4.3 Mm & 43 & 100 & 95  \\
6.2 Mm & 39 & 97 & 83  \\
9.2 Mm & 53 & 71 & 66  \\
\hline
\end{tabular}
\caption{Summary of parameters of inversions for horizontal flow components at various depths. I give level of noise in the results and root-mean-square values of both horizontal components. The values represent the average over nine 24-hour measurements. }
\label{tab:tomography_parameters}
\end{table}

\begin{table}
\begin{tabular}{c|cccccc}
& 0 Mm & 1.9 Mm & 2.9 Mm & 4.3 Mm & 6.2 Mm & 9.2 Mm\\
\hline
0 Mm & $1.00$ & $0.45$ & $0.46$ & $0.64$ & $0.39$ & $-0.10$ \\ 
1.9 Mm & $0.50$ & $1.00$ & $0.59$ & $0.48$ & $0.11$ & $-0.33$ \\ 
2.9 Mm & $0.40$ & $0.64$ & $1.00$ & $0.72$ & $0.15$ & $-0.25$ \\ 
4.3 Mm & $0.62$ & $0.53$ & $0.73$ & $1.00$ & $0.44$ & $-0.24$ \\ 
6.2 Mm & $0.28$ & $0.12$ & $0.01$ & $0.34$ & $1.00$ & $0.42$ \\ 
9.2 Mm & $-0.25$ & $-0.35$ & $-0.39$ & $-0.38$ & $0.31$ & $1.00$ \\ 
\hline
\end{tabular}
\caption{Values of mutual Pearson's correlation coefficient for estimates of the horizontal flow components for six investigated depths. Above the diagonal of the table the correlation coefficients for $v_x$ are given, below diagonal then the correlation coefficients for $v_y$ estimates.}
\label{tab:tomography_correlation}
\end{table}

\end{document}